\documentclass[12pt]{article}

\usepackage{amstext,amsmath,amssymb,amsfonts,bbm}
\usepackage[latin1]{inputenc}
\usepackage{hyperref}
\usepackage{amsthm}
\usepackage{color}
\usepackage{graphicx}
 
\newtheorem{definition}{Definition}
\newtheorem{theorem}{Theorem}
\newtheorem{lemma}{Lemma}
 
\newcommand{\cA}{{\mathcal A}}
\newcommand{\cG}{{\mathcal G}}
\newcommand{\cT}{{\mathcal T}}
\newcommand{\cL}{{\mathcal L}}
\newcommand{\cF}{{\mathcal F}}
\newcommand{\cR}{{\mathcal R}}
\newcommand{\Tr}{\mathrm{Tr}}

\newcommand{\be}{\begin{equation}}
\newcommand{\ee}{\end{equation}}
  
\newcommand{\bea}{\begin{eqnarray}}
\newcommand{\eea}{\end{eqnarray}}

\newcommand{\prf}{{\noindent \bf Proof\; \; }}

\begin{document}
\begin{flushright}
 
\end{flushright}

\vspace{20pt}

\begin{center}

{\Large\bf 
Double Scaling in Tensor Models\\\vskip.3cm {with a Quartic Interaction}}
\vspace{20pt}

St\'ephane Dartois${}^{a,b}$
\footnote{e-mail: stephane.dartois@ens-lyon.fr}, 
Razvan Gurau{}$^{c,d}$
\footnote{e-mail: rgurau@cpht.polytechnique.fr} 
and Vincent Rivasseau${}^{b,d}$
\footnote{e-mail: vincent.rivasseau@th.u-psud.fr}

\vspace{10pt}

\begin{abstract}
\noindent In this paper we identify and analyze in detail the subleading contributions in the $1/N$ expansion of 
random tensors, in the simple case of a quartically interacting model. The leading order
for this $1/N$ expansion is made of graphs, called \emph{melons}, which are dual to particular 
triangulations of the $D$-dimensional sphere, closely related to the ``stacked'' triangulations. 
For $D<6$ the subleading behavior is governed by
a larger family of graphs, hereafter called \emph{cherry trees}, which are also 
dual to the $D$-dimensional sphere. They can be resummed explicitly through a double scaling limit.
In sharp contrast with random matrix models, this double scaling limit is stable.
Apart from its unexpected upper critical dimension 6, it displays a singularity at fixed distance from the origin
and is clearly the first step in a richer set of yet to be discovered multi-scaling limits.
\end{abstract}
\end{center}


\setcounter{footnote}{0}
\setcounter{lemma}{0}
\setcounter{theorem}{0}

\section{Introduction}

Tensor models  
are the natural generalization of matrix models
\cite{Mehta,matrix},
implementing in a consistent way statistical sums over random triangulations. 
The Feynman graphs of matrix models, also called ribbon graphs, are dual to \emph{maps}, hence to discretizations of Riemann surfaces.
Early rank-$D$ tensor models were introduced in order to generalize this remarkable property to higher dimensions $D\ge 3$ \cite{oldgft1,oldgft2,mmgravity}.
Until recently they were difficult to handle analytically, hence were  mostly studied through numerical simulations. Group field theory \cite{Boul,freidel,Oriti1} 
is a related approach
which incorporates in a common quantum field theory framework the vertex structure of early tensor models and the spin-foam amplitudes of covariant loop 
quantum gravity \cite{LQG1,LQG2,LQG3}. However early tensor models and GFTs had two major shortcomings: first they did not triangulate pseudo-manifolds
with well defined $D$-homology; and even more importantly they lacked an analog of the powerful $1/N$ expansion of matrix models \cite{'tHooft:1973jz}. 
This expansion structures the ordinary Feynman perturbative expansion according to the genus of the triangulated Riemann surface. It is led by the family 
of regular planar graphs, 
which can be counted exactly \cite{Tutte,Brezin:1977sv}.  It is this matrix $1/N$ expansion which gave access to the single \cite{mm,Kazakov:1985ds} 
and double scaling 
limits \cite{BKaz,DS,GM} of matrix models, hence to the \emph{functional integral quantization of two dimensional gravity}. 
The perturbative expansion of string theory
is also structured similarly by the genus of the string world sheet.

The search for improved GFTs free of these two shortcomings led to the 
introduction of \emph{colored} group field theory \cite{color,sefu2,review}. The corresponding Feynman graphs have edges of 
$D+1$ different colors  meeting at vertices of degree $D+1$. This leads to a well-defined $D$-homology for the dual triangulations \cite{lost,pezzana,FG}.
An associated tensorial $1/N$ expansion was soon discovered for such colored models \cite{expansion1, expansion2, expansion3}, 
providing the missing tool to structure their perturbative expansion into an interesting hierarchy. 
This tensorial $1/N$ expansion is indexed by an integer called the \emph{degree}. This degree
is no longer a topological invariant, as was the case for matrices, but the sum over genera of \emph{jackets}.
Jackets are ribbon graphs embedded in the colored tensor graphs which provide $D!/2$ global Heegaard decompositions 
of the triangulation \cite{ryan}.
 
The leading order of the tensorial $1/N$ expansion is by now well understood, and shows the same structure in any dimension $D\ge 3$. 
It is governed by a family of leading graphs, called \emph{melons}, closely related to ``stacked'' triangulations of the
$D$ dimensional-sphere. The corresponding single scaling critical point \cite{critical} leads to a continuous phase of branched polymers 
\cite{melbp}.       

Coloring was soon recognized also as essential to formulate a general theory of \emph{unsymmetrized} random tensors \cite{univ,uncoloring}. This theory
is characterized by its invariance under the tensorial product of $D$ independent copies of the $U(N)$ group. This invariance should be considered as an abstract 
generalization of the \emph{locality} of interactions in ordinary field theory. Natural tensor interactions are polynomial invariants in the tensor coefficients,
associated to bipartite colored graphs. They generalize the cyclic trace interactions of matrix models. In such so called ``uncolored" models,
colors reduce to a book-keeping device which tracks the position of the indices of the tensor.
The tensor track program \cite{Rivasseau:2011hm,Rivasseau:2012yp} proposes a systematic investigation of such tensor models and of their related 
tensorial group field theories in order to develop the corresponding functional integral quantization of general relativity in dimension higher 
than two\footnote{Recall that the two natural parameters of tensor models, namely the coupling constant and the size $N$, can be interpreted as 
discrete versions of the cosmological constant and of Newton's constant.}.

Slightly breaking the $U(N)$ invariance of matrix models leads to renormalizable non-commutative quantum field theories \cite{Rivasseau:2007ab},
such as the Grosse-Wulkenhaar (GW) model \cite{Grosse:2004yu} whose planar sector has recently been beautifully solved \cite{GWplanar}. The key difference
between random matrix models and NCQFT's lies in the modification of the propagator. The GW propagator breaks the $U(N)$ invariance of the 
theory in the infrared (small $N$) regime, but is asymptotically invariant in the ultraviolet regime (large $N$); hence it
launches a renormalization group flow between these two regimes. Power counting 
in such models is entirely governed by the underlying $1/N$ expansion (divergent graphs are the regular planar graphs),
hence the corresponding  \emph{matrix} renormalization group can be considered the continuous version of the $1/N$ expansion. It is remarkable
that this matrix renormalization group flow is asymptotically safe \cite{DGMR}.

In a completely analogous manner, renormalizable quantum field theories of the tensorial type have been defined 
through a soft breaking of the tensorial invariance of their propagator \cite{BenGeloun:2011rc,Geloun:2012bz,Geloun:2013saa}.
Renormalization is again a continuous version of the $1/N$ expansion and the divergent graphs are the melons.
The corresponding \emph{tensorial} renormalization group flow generically displays the fundamental
physical property of asymptotic freedom \cite{BenGeloun:2012yk,BenGeloun:2012pu,jbgelon}.
Renormalizable GFTs which include a "Boulatov-type" propagator
have also been recently defined and studied \cite{cor1,VTD,cor2}. They incorporate as additional data 
discretized Lie-group valued connections over the dual triangulated pseudo-manifold, allowing to consider 
other metric data than the graph distance on this pseudo-manifold. Remarkably they are also asymptotically free \cite{Samary:2013xla,Carrozza}.

Other recent developments in this rapidly evolving field include extension of the $1/N$ expansion to more general tensor models \cite{expansion4,expansion5}, 
bounds on the number of spherical triangulations \cite{rarespheres},
investigation of the Schwinger-Dyson equations of tensor models  \cite{sdequations,sdequations1,sdequations2,sdequations3}  and
of the Hopf algebraic structure of tensorial renormalization \cite{Hopftens} and some preliminary but promising applications to
statistical mechanics in presence of random geometry \cite{IsingD,EDT,spinglass,Bonzom:2012sz,Bonzom:2012qx}.

In another remarkable recent progress a rigorous non-perturbative treatment of the $1/N$ expansion 
was given using an \emph{improved loop vertex expansion} (LVE) \cite{Gurau}\footnote{The  improvement, which consists in replacing resolvents by their
parametric representation, allows to factorize traces along faces of the associated graphs.}.
The LVE  \cite{LVE1,LVE21,LVE2} combines the intermediate field representation with
the tree formula \cite{BK,AR} and replica trick of constructive field theory \cite{GlimmJaffe,Rivasseau:1991ub}. In contrast 
with ordinary perturbative expansion it is absolutely convergent, mathematically defining the correlation functions of the model as the
Borel sum \cite{NevSok}  of their perturbative series on the stable side of the coupling constant. 

The LVE is in fact better adapted to the analysis of tensor models than to matrix models. Indeed it
naturally allows a systematic computation of the tensorial $1/N$ expansion. In fact it reduces rank-D tensor integration to 
integration over D random matrices, themselves made of replicas (one per vertex of the LVE). 
The leading terms, the melons, correspond to the trees of the LVE with intermediate matrix $\sigma$ fields
all put to zero. The  sub-leading effects in $N$ correspond to adding a finite number of loops
made of Wick-contractions of these fields\footnote{Remark that in this representation the LVE trees form a kind of 
``primitive space-time" on which the intermediate matrix fields live as fluctuation fields, generating the subleading effects. 
This picture is quite appealing from the point of view of geometrogenesis \cite{Konopka:2006hu,Oriti:2013jga}.}.
A mixed expansion can be written in both the coupling $\lambda$
and $N$, which allows a systematic investigation of these sub-leading effects. The situation is not identical to the random matrix case,
in which all the planar terms are of the same order in 1/N as the LVE trees.

In this paper we use the tensorial LVE to find the equivalent of matrix double scaling for tensor models. Prior to this work, a simpler
version of ``matrix-like" double scaling limit for tensors was investigated in \cite{doubletens} and the first term beyond melons in the 
$1/N$ expansion has been identified in \cite{KOR}. But during the course of our investigation we had to answer new questions among which:

\begin{itemize}
 \item How to identify the right scaling rate between $N$ and $ \lambda$ and the class of graphs 
 contributing to the double scaling? These questions are intimately intertwined.
 \item Is such a double expansion summable? The answer is negative in the case of matrix models 
 but turns out to be positive for tensors in $D<6$. 
\end{itemize}

Our main results 
are contained in Theorem \ref{thm:ampli} and \ref{maintheo} in section \ref{sec:GraphandTheo}, which in $D<6$ separates the next-to-leading family of \emph{cherry-trees} from the rest,
and in Theorem \ref{doublelimtheo} in section \ref{sec:Double}, which performs the proper double-scaling limit by summing over cherry trees. 
The picture emerging from our analysis is that in dimension $D<6$ the double scaling has the same susceptibility 1/2 as the single scaling, hence 
it is possible that it leads again to a branched polymer phase. But the important points are elsewhere. 
First at $D=6$ or above, the double scaling incorporates larger families of graphs which may drive 
the system to a different phase (which remains to investigate in a future study). Second, and more important, the double 
scaling limit for $D<6$ seems to lead to further multiple scalings which may be probed by similar techniques. This is also 
left for future study, but we can speculate that the ensuing continuous phase obtained after a finite number of such 
steps in dimension $D=3$ or $4$ is different from branched polymers.

The reader could wonder whether there is a discrepancy between our analysis and the one of \cite{KOR}, which finds a critical susceptibility $3/2$
for the next-to-leading order of the tensorial $1/N$ expansion. In fact there is no discrepancy, since double scaling resums an infinite family 
of contributions, the first one of which corresponds to the one analyzed in \cite{KOR}.  Although the model in \cite{KOR} is colored, hence not identical to ours,
we have checked that the \emph{first order cherry trees} of our model also give the critical susceptibility $3/2$ found in \cite{KOR}.

During preparation of this paper another parallel study lead to results similar to ours 
for the double scaling of \emph{colored} tensor models \cite{GurauSchaeffer}. 
The existence of the upper dimension 6 was found independently and on the same day by the two approaches.
On one hand, the colored model treats a more general category of graphs than the uncolored quartic one, hence our analysis could 
be recovered as a particular case of \cite{GurauSchaeffer}. On the other hand the methods are very different, as well as the emphasis,
since the colored Bosonic model does not correspond to a stable interaction hence cannot be treated directly by the LVE.
Hence the two works will be published separately.

This paper is structured as follow: in section \ref{sec:Matrix} results about $1/N$ expansion and the double scaling limit of matrix are recalled. 
In section \ref{sec:color} 
we introduce the basics of tensor models as well as the constructive QFT ideas that led to this work. In section \ref{sec:GraphandTheo} we introduce useful 
graphs definitions, manipulations and state our main results which identify the leading graphs -called \textsl{cherry trees}- and 
clarifies in which sense they dominate. Sections \ref{sec:examples} and \ref{sec:proof} give the proofs of our results, 
and the double scaling limit of the model (Theorem \ref{doublelimtheo}) is established in section \ref{sec:Double}.

\section{Matrix Models}
\label{sec:Matrix}

We start by briefly recalling the classical results concerning the continuum limit and double scaling 
in matrix models.

\bigskip

\noindent{\bf Matrix Single Scaling.} The non-Hermitian matrix model with a quartic interaction is the probability 
measure 
\be  \label{eqmat1}
 d\nu(M) = \frac{1}{Z(\lambda)}    \Bigg( \prod_{i,j=1}^N \frac{ d M_{ij} d \bar M_{ij}}{2\pi \imath } \Bigg) \;
e^{- N \bigl[ \Tr (MM^{\dagger} ) + \lambda \Tr(MM^{\dagger} MM^{\dagger}) \bigr] } \; ,
\ee 
where the normalization constant $Z(\lambda)$ is known as the partition function. The free energy 
of the model $F(\lambda) = \ln Z(\lambda)$ as well as any other correlation function (like for instance
the two point function)
\be \label{eqmat2}
 G_2(\lambda) = \frac{1}{Z(\lambda)}  \int  \Bigg( \prod_{i,j} N \frac{ d M_{ij} d \bar M_{ij}}{2\pi \imath } \Bigg) \;
\Bigl( \frac{1}{N} \Tr(MM^{\dagger}) \Bigr)  e^{- N \bigl[ \Tr (MM^{\dagger} ) + \lambda \Tr(MM^{\dagger} MM^{\dagger}) \bigr] } \; ,
\ee
admit an expansion in Feynman graphs. The graphs contributing to $F(\lambda)$ are ribbon graphs with quartic vertices and no external legs, 
while the graphs contributing to the two point function $G_2(\lambda)$ are ribbon graphs with quartic vertices and a marked edge. In 
the large $N$ limit only planar graphs survive and the planar two point function is
\bea
 G_{2, \text{planar} }(\lambda) = \frac{ -1 - 36\lambda + (1+24 \lambda)^{3/2}   }{ 216 \lambda^2 } 
 = \sum_{n\ge 0} \frac{2}{(n+2)} \frac{3^n}{(n+1)} \binom{2n}{n} (-2\lambda)^n \; ,
\eea 
reproducing Tutte's counting of planar quadrangulations with a marked, oriented edge. The two point function
exhibits a critical behavior at $\lambda_c=-\frac{1}{24}$, 
$G_{2, \text{planar } }(\lambda) \sim ( \lambda - \lambda_c )^{1-\gamma}$, with critical exponent
exponent\footnote{This exponent is known as the string susceptibility exponent, or
as the entropy exponent: it characterizes asymptotically the number of planar graphs with a fixed number
of vertices.}  $\gamma = -\frac{1}{2}$, corresponding to pure gravity in $D=2$ \cite{matrix}.

\bigskip

\noindent{\bf Matrix Double Scaling.} Subsequent orders in the $1/N$ expansion of matrix models are accessed in the double scaling limit \cite{BKaz,DS,GM}.
The $1/N$ expansion of the two point function is indexed by the genera of the surfaces triangulated by the Feynman graphs of the model. 
One can thus write this expansion as:
\be
G_2[ \lambda, N]= \sum_{h=0}^\infty N^{ -2h} G_{2,h} (\lambda) \; ,
\ee
where $G_{2,h}(\lambda)$ is given by the sum over quadrangulations with a marked oriented edge and fixed genus $h$ \cite{Schaeffer},
\be
G_{2,h} (\lambda) \simeq \sum_{n} n^{ \frac{5}{2} (h-1) }
\Bigl( \frac{\lambda}{\lambda_c}\Bigr)^n\simeq a_h (\lambda-\lambda_c)^{\frac{3}{2} - \frac{5}{2}h} \; .
\ee
The double scaling limit consists in taking the limit $N\rightarrow \infty$ and $\lambda\rightarrow \lambda_c$ in a correlated way. 
In fact one remarks that all $ G_{2,h} $ have the same critical point $\lambda_c$, thus their contribution is enhanced when 
$\lambda\rightarrow \lambda_c$. Defining
\be
\kappa^{-1}=N^{5/4}(\lambda-\lambda_c) \; ,
\ee
we take the limit $N \to \infty, \lambda \to \lambda_c$ keeping $\kappa$ fixed and we obtain the double scaling limit of the two point function
\be
G_2 [\kappa]= \sum_{h\ge 0}\kappa^{2h}a_h .
\ee
Unfortunately, as remarked first in \cite{Daviddouble}, this expansion in $\kappa$ is not summable. 
First, the $a_h$'s are all positive. This comes from the fact that the critical point $\lambda_c = -\frac{1}{24}$ is on the 
negative real axis, hence it corresponds to an unstable potential $e^{+N|\lambda_c| \Tr (MM^{\dagger}MM^{\dagger})}$.
Second, the $a_h$'s grow too fast with $h$, $a_h \simeq (2h)!$. This comes from the fact that in the double scaling limit
one ends up summing over all the Feynman graphs at once and one encounters the usual problem that the perturbative
expansion generates too many graphs.

\section{Tensor Models} \label{sec:color}

The general framework of invariant tensor models is presented in detail in 
\cite{uncoloring,univ,Gurau}. We summarize here the main points required 
for this paper. We deal with rank $D$ covariant tensors ${\mathbb T}_{n^1\dots n^D}$, with $n^1, n^2,\dots n^D \in \{ 1,\dots N\}$
having \emph{no symmetry} under permutation of their indices. The complex conjugate tensor, $\bar {\mathbb T}_{  n^1 \dots n^D }$,
is a rank $D$ contravariant tensor. They transform under the
{\it external} tensor product of $D$ fundamental representations of the unitary group $U(N)$
\bea
&& {\mathbb T}_{a^1\dots a^D} = \sum_{n^1\dots n^D}U^{(1)}_{a^1n^1}\dots U^{(D)}_{a^Dn^D} {\mathbb T}_{n^1\dots n^D}  \; ,\crcr
&&  \bar {\mathbb T}_{ \bar a^1\dots  \bar a^D} = \sum_{ \bar n^1\dots \bar n^D}
\bar U^{(1)}_{\bar a^1 \bar n^1 }\dots \bar U^{(D)}_{  \bar a^D \bar n^D} \bar {\mathbb T}_{ \bar n^1\dots \bar n^D}  \; ,
\eea
where the indices of the complex conjugated tensor are denoted conventionally with a bar.
Each unitary group acts separately on an index: the unitary operators $U^{(1)},\dots U^{(D)}$ 
are all {\it independent} (one can even consider unitary groups of different sizes $U(N_i)$, one for each index $n^i$).
We use the shorthand notation $\vec n$ for the $D$-tuple of integers $(n^1, \dots n^D)$, and we consider $D\ge 3$.

Any invariant polynomial in the tensor entries can be expressed in terms of the \emph{trace invariants}
built by contracting in all possible ways pairs of covariant and contravariant indices in a product 
of tensor entries. Of particular importance in the sequel are the quartic ``melonic'' invariants
\be
\sum_{n  \bar n  }
    \mathbb{T}_{\vec n }  \bar { \mathbb{T}}_{ \vec {\bar m} }
\mathbb{T}_{ \vec m }  \bar { \mathbb{T}}_{ \vec {\bar n} } \;  \delta_{n^i\bar m^i} \delta_{m^i \bar n^i}
\prod_{j\neq i}  \delta_{n^j \bar n^j} \delta_{m^j \bar m^j} \; .
\ee
They are built from four tensor entries, two ${\mathbb T}$ and two $ \bar{\mathbb T}$, such that all indices except the one in the position $i$ 
are contracted between a pair  $({\mathbb T}, \bar{\mathbb T} )$ while the indices in the position $i$ are contracted between the pairs. 

\begin{figure}[ht]
   \begin{center}
 \includegraphics[width=8cm]{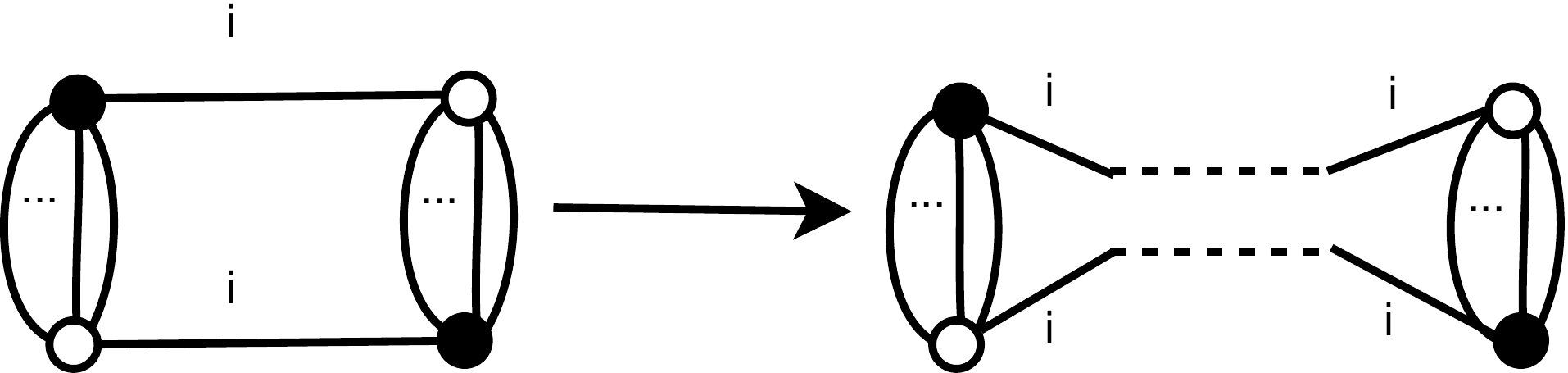}
 \caption{The Quartic Melonic Interactions}
  \label{quartic}
   \end{center}
\end{figure}

The trace invariants can be represented as bipartite closed $D$-colored graphs. The graph associated to an invariant
is obtained as follows:
\begin{itemize}
 \item we represent every ${\mathbb T}_{ n^1\dots n^D}$ (respectively $ \bar {\mathbb T}_{ \bar n^1\dots \bar n^D}$) by a white vertex 
      $v$ (respectively a black vertex $\bar v$)
 \item we represent by an edge of color $i$ the contraction of an index $n^i$ on ${\mathbb T}_{ n^1\dots n^D}$ with an index $\bar n^i$
     of $ \bar{\mathbb T}_{ \bar n^1\dots \bar n^D}$.
\end{itemize}
The graph of a melonic quartic invariant is represented in Fig. \ref{quartic}.
Colored graphs are dual to $D$-dimensional abstract simplicial pseudo-manifolds \cite{color,lost,pezzana,FG}, hence the tensor models
encode a theory of random triangulated higher dimensional spaces.

The quartically perturbed Gaussian tensor measure with which we will deal in this paper (see \cite{Gurau}) is the simplest 
interacting tensor model, with measure
\bea\label{eq:measure}
&& d\mu =  \frac{1}{Z(\lambda,N)}  \Big{(}\prod_{ \vec n } N^{D-1} \frac{d \mathbb{T}_{\vec n} d\bar {\mathbb{T} }_{ {\vec n } } } { 2 \pi \imath } \Big{)} \;
   e^{-N^{D-1}  S^{(4)}( \mathbb{T} ,\bar { \mathbb{T} } ) } \; , \\
&&  S^{(4)}( \mathbb{T},\bar { \mathbb{T} } ) =   \sum_{\vec n } \mathbb{T}_{\vec n } \delta_{ \vec n  \bar {\vec n}   } \bar { \mathbb{T} }_{ \bar {\vec n} }
    + \lambda \sum_{i=1}^D \sum_{n  \bar n  }
    \mathbb{T}_{\vec n }  \bar { \mathbb{T}}_{ \vec {\bar m} }
\mathbb{T}_{ \vec m }  \bar { \mathbb{T}}_{ \vec {\bar n} } \;  \delta_{n^i\bar m^i} \delta_{m^i \bar n^i}
\prod_{j\neq i}  \delta_{n^j \bar n^j} \delta_{m^j \bar m^j} \; , \nonumber
\eea 
with $Z(\lambda,N) $ some normalization constant. 

Some explanations here are in order. The $D$ different quartic interactions correspond to those of Fig. \ref{quartic}, namely to the
leading quartic melonic stable interactions at tensor rank (i.e. dimension) $D$. The scaling in equations \eqref{eq:measure} is {\it the only} 
scaling leading to a large $N$ limit. This is a nontrivial statement and the reader should consult \cite{univ}
for its proof.

The melonic interactions turn out to be the easiest ones to decompose according to intermediate fields. Higher order stable melonic (or even submelonic) interactions
can in principle be treated as well by the same method but require more intermediate fields \cite{LVE2}. In principle
each of the $D$ interactions can have its own different coupling constant $\lambda_i$, but in this paper
we shall treat only the symmetric case in which all $\lambda_i$ are equal to $\lambda$.

The two point function of the model  
\bea
 \cG_2 =  \frac{1}{Z(\lambda,N)}  \int \Big{(}\prod_{ \vec n } N^{D-1} \frac{d \mathbb{T}_{\vec n} d\bar {\mathbb{T} }_{ {\vec n } } } { 2 \pi \imath } \Big{)} \;
   \Bigl( \frac{1}{N} \sum_{n } \mathbb{T}_{\vec n }  \bar { \mathbb{T}}_{ \vec {\bar n} }  \Bigr)  \prod_i \delta_{n^i\bar n^i}
      e^{-N^{D-1}  S^{(4)}( \mathbb{T} ,\bar { \mathbb{T} } ) } \; ,
\eea 
is well defined on the ``constructive side'' $\Re \lambda >0$, on which its (alternated) perturbative series is 
Borel summable \cite{Gurau}. Performing the $1/N$ expansion of the model, either in perturbation theory 
or in the constructive LVE expansion \cite{univ,Gurau} one obtains that the leading order is given by melonic graphs (corresponding to
the trees in the LVE expansion)
\bea
 \lim_{N\to \infty} \cG_2 = \cG_{2,\text{melon} } = \frac{-1+ \sqrt{1+8D\lambda} } {4D\lambda} \; .
\eea
This corresponds to a critical constant $\lambda_c = -\frac{1}{8D}$ and a susceptibility exponent $\gamma = \frac{1}{2}$.
Note that, as in the matrix models case, the singularity of the leading order in the $1/N$ expansion is on the negative real axis,
corresponding to the sum with only positive signs over the class of melonic $D$ dimensional triangulations with a marked $D-1$ simplex.

The full perturbative expansion of the two point function can be reorganized as a series in $1/N$
\bea
 \cG_2 = \cG_{2,\text{melon} } + \sum_{h\ge 1} \frac{1}{N^{h}} \cG_{2,h} \; .
\eea 

The $1/N$ series for tensor models is considerably more complicated than for matrix models: only some values of $h$ are allowed,
the families $\cG_{2,h}$ are complicated, and so on. The idea of the double scaling is the following. When sending $\lambda$ to 
$\lambda_c$, the melons become critical. One can then restrict to graphs $\bar \cG$ with no melonic subgraphs and sum, for every 
such graph, the family of graphs obtained by arbitrary insertions of melons. 
This will lead to an expansion of the two point function 
\bea
 \cG_2 & = & \cG_{2,\text{melon} } + \sum_{\bar G} \frac{1}{N^{h(\bar G )}} \frac{1}{(\lambda - \lambda_c)^{e(\bar G)}} \crcr
 & =& \cG_{2,\text{melon} } + \sum_{e\ge 1} \sum_{\bar G, e(\bar G) = e} 
   \Bigl( \frac{1}{N^{ h(\bar G)/e(\bar G)} (\lambda - \lambda_c) }  \Bigr)^{e}
\eea 
where $h(\bar G)$ is the scaling with $N$ of $\bar G$, $e(\bar G)$ counts the number of places where we can insert melons
in $\bar G$. We now select the family of graphs for which $ h(\bar G)/e(\bar G)$ is minimal.
Denoting this minimal value $\alpha$, the two point function becomes 
\bea\label{eq:doublescale}
 \cG_2 &=& \cG_{2,\text{melon} } + \sum_{e\ge 1} \Bigl[ \sum_{\bar G, e(\bar G) = e,\alpha (\bar G ) = \alpha} 
   \Bigl( \frac{1}{N^{\alpha} (\lambda - \lambda_c) }  \Bigr)^{e} \crcr
      && \qquad \qquad \qquad +   
   \sum_{\bar G, e(\bar G) = e, h (\bar G ) /e(\bar G) > \alpha} 
   \Bigl( \frac{1}{ N^{ h(\bar G)/e(\bar G) -\alpha } N^{\alpha} (\lambda - \lambda_c) }  \Bigr)^{e}
   \Bigr] \; .
\eea 
When tuning $ \lambda $ to $\lambda_c$ while keeping $ N^{\alpha} (\lambda - \lambda_c)   $ fixed 
the last sum cancels and all the terms in the first sum admit a well defined limit. The expansion obtained
in this limit is an expansion in a new double scaled parameter $x = N^{\alpha} (\lambda - \lambda_c)$.
The rest of this paper is dedicated to establishing \eqref{eq:doublescale}.
As a matter of notations, in the rest of this paper we set $z = -2\lambda$.

A very convenient way to organize the sum over graphs $\bar G$ is to use the constructive LVE representation of 
the quartic tensor model introduced in \cite{Gurau}. 

\subsection{Constructive QFT}

The main idea of constructive theory is to reorganize perturbation
theory around the (classical) notion of trees, rather than around the (quantum) notion of (Feynman) graphs.
Like graphs, trees have a main advantage: they show connectivity, hence allow to compute
logarithms of functional integrals, the fundamental step in QFT. But
since trees are much fewer than graphs, they can perform this computation 
in a convergent way.

\bigskip

\noindent{\bf Weights.} The constructive reorganization of perturbation theory uses combinatoric weights defined for any pair $(G,T)$
made of a connected graph $G$ and a spanning tree $T \subset G$ \cite{RZTW}.
The simplest definition of these weights is through \emph{Hepp sectors}.
A Hepp sector \cite{Hepp} is a total ordering of the edges of a Feynman graph.
Consider a connected graph $G$ with $V$ vertices, $E$ edges and $L = E-V +1$ loop edges (independent cycles).
Any spanning tree $T$ has $V$ vertices, $V-1$ edges and induces a partition of $G$ into the tree $T$ and the set $\cL$ of the $L$ loop edges. \
For any Hepp sector $\sigma \in {\mathfrak S}_E$ of $G$ there is a single associated minimal spanning tree of $G$, denoted $T(\sigma)$, obtained by 
Kruskal ``greedy " algorithm \cite{kruskal}, that keeps recursively 
the edges with the smallest possible indices in $\sigma$. 
Defining $N(G,T) = \# \{ \sigma , T(\sigma) =T \} $, the weights $w(G,T)$ are simply
the percentage of sectors  $\sigma$ of $G$ for which $T(\sigma) =T$:
\be  w(G,T) = \frac{N(G,T)}{E!} .
\label{favoriteweights}
\ee
Hence these weights are rational numbers which for any graph $G$ define a probability measure on the set of 
its spanning trees:
\be \sum_{T\subset G}  w(G,T) =1 . \label{bary}
\ee
They are symmetric in terms of relabeling of the vertices $\{v_1, \cdots v_n \}$ of the graph.
A main property is their positive type integral representation 
\begin{lemma}[\cite{RZTW}]
\bea
w(G,T) =   \int_0^1 \prod_{\ell \in T} du_\ell \prod_{\ell \in G\setminus T} w^T_{\ell}(\{u\})  \label{favoriteweights1} 
\eea 
where $w^T_{\ell}(\{u\})$ is the infimum over the $u_{\ell'}$ parameters 
over the lines $\ell'$ forming the {\it unique} path $P^T_\ell$ in $T$ joining the ends of $\ell$.
\end{lemma}

Remark that $P^T_\ell$, hence also $w^T_{\ell}(\{u\}) $, only depends on the end vertices $i$ and $j$ of $\ell$.
The representation \eqref{favoriteweights1} is called positive type 
because the real symmetric $V$ by $V$ matrix $W^T_{i,j}(u) =  w^T_{ij}(\{u\})$ 
(completed by $W^T_{i,i}(u) =1$ on the diagonal) is of \emph{positive type} for any $u_\ell \in [0,1]^{\vert T\vert }$ \cite{BK,AR}.
This is what allows to bound amplitudes of the LVE representation and prove its absolute convergence \cite{LVE1,Gurau}.
Uniformly distributed weights $w(G,T) = 1/\chi (G)$, where $\chi (G)$
is the complexity of $G $ i.e. the number of its spanning trees, cannot be written in general
as integrals such as \eqref{favoriteweights1} with positive-type matrices $w^T_\ell$, hence they do not lead
to a convergent repacking of perturbation theory.

Note that each edge has an associated continuous variable: either a $u$ variable for a tree edge or a $w$ variable for a loop edge.
The parameters of the tree edges in the unique path connecting the end points of a loop edge are larger than the parameter 
of the loop edge. 

\bigskip

\noindent{\bf The Loop Vertex Expansion.} Reorganizing perturbation theory according to these weights allows one  
to decompose the Feynman amplitudes of any connected QFT quantity $S$ in terms of the spanning trees 
inside each Feynman graph \cite{RZTW}. One starts from the functional integral (in our case the tensor integral)
and first performs an expansion indexed by trees 
\be
S=  \sum_T  \cA(T)   \;,  \quad   \quad  \vert    \sum_{T} \cA(T) \vert < \infty  \;.
\ee
This expansion is not the usual perturbative expansion and it has the main advantage that the sum over trees converges. 
Subsequently the contribution of each tree can be re-expanded as a sum over all graphs in which the tree is spanning
\be
\cA(T) =  \sum_{G \supset T}  w(G,T) \cA(G) \Rightarrow
 S = \sum_T  \cA(T)  =  \sum_{T } \sum_{G; G \supset T}  w(G,T) \cA(G) \; . 
\ee
The subtle point is that one cannot naively exchange the sum over trees with the sum over graphs: the sum over graphs is not a summable
series. However the sum over graphs is Borel summable, and the series indexed by the trees
is its Borel sum. The definition of $S$ as a sum over trees (and not over graphs) is in fact its {\it correct} definition.

For Bosonic QFTs with stable interactions one must add a further twist. It turns out that the most convenient 
representation of $S$ as a sum over trees is obtained \emph{in terms of the intermediate field representation}
\cite{LVE1}
\footnote{For Fermionic theories the use of the intermediate field representation is not required \cite{AR}, as
 Borel summability is not necessary as long as cutoffs are finite.}.
The trees of the intermediate field representation are \emph{plane trees} 
because the vertices of the intermediate field representation are cycles (from the point of view of the original model)
made out of an arbitrary number of the ordinary propagators (hence the name "Loop Vertex Expansion").
The LVE requires decomposing the interaction into three body interactions, hence works best for quartic perturbations,
but can be extended to more general stable perturbations \cite{LVE2}.

Let $\cT$ a plane  tree with $n$ vertices labeled $1, 2,\dots n$. Consider a parameter $u_\ell \in [0,1]$ for each edge $\ell \in \cT$ and 
a collection of $n$ replica fields $\sigma = \{\sigma^1, \cdots \sigma^n \}$, one for each vertex of the tree.
Positivity of the matrix $W$ allows one to define, for every value of the $u_\ell$ parameters,
the unique Gaussian  measure $ \mu^{\cT} (u)$  over
the set of random fields $\sigma$ of covariance
\be
\int d \mu^{\cT} (u)  \; \; \sigma^{i}  \sigma^{j}  = w_{ij}^{\cT}(u ) .
\ee

The LVE representation expresses $\cA(\cT)$ as
\be
\cA(\cT)  = \int d \mu^{\cT} (u)  \;  \; \;   \Tr \Big[  \prod^{\rightarrow}_{q \in C(\cT)}  R (\sigma^{i(q)})  \Big]
\ee
where $q$ runs over the set $C(\cT)$ of the $2(n-1)$ \emph{corners} of the plane tree, $\rightarrow$ 
means that the product is ordered along the Dyck path turning around the plane tree and 
$i(q)$ is the vertex reached at step $q$ in this path.
The resolvent  $R(\sigma)$ is defined as 
$R(\sigma) = [1 + i H (\sigma ) ]^{-1}$ where $H( \sigma)$ is the Hermitian operator
of the type $C^{1/2}\sigma C^{1/2}$   where $C$ is the covariance of the initial ordinary Bosonic field.
Absolute convergence of the sum over trees follows from the simple fact that  by Hermiticity of $H( \sigma) $
the norm of each such resolvent is bounded by 1 \cite{LVE1,LVE2}.

The LVE was initially introduced to prove that the $N$-dependence of the constructive bounds for matrix QFTs reproduces 
the perturbative one \cite{LVE1}. In the matrix case however, the LVE does not by itself compute the $1/N$ expansion. 
Indeed the LVE is naturally organized into trees plus loop corrections on these trees. Planar loop corrections to the LVE 
tree terms are not smaller than the contribution of the trees, but of the same order in $1/N$. This is not surprising: we know that 
the dominant term in the matrix $1/N$ expansion are planar graphs, which are not just simply 
trees\footnote{Planar graphs are in one-to-one correspondence with \emph{well-labeled trees} via
 the Cori-Vauquelin-Schaeffer bijection \cite{Schaeffer}.}.

The LVE works in fact better for tensors than for matrices: its leading (tree) term directly
computes the leading $1/N$ term, at least in the case of the simple quartic model \eqref{eq:measure}. 
\begin{figure}[ht]
   \begin{center}
 \includegraphics[width=12cm]{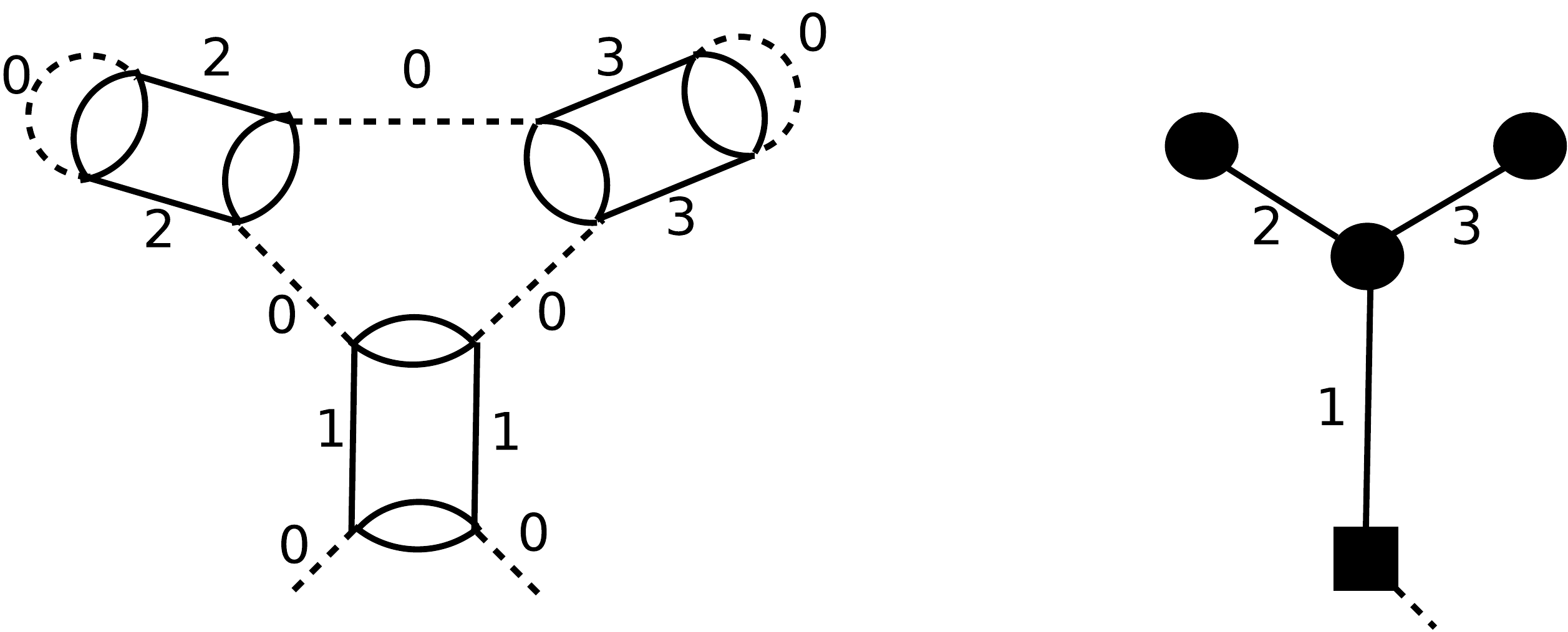}
 \caption{A LVE tree in the direct (left) and intermediate field (right) representation for a quartic interaction.}
  \label{bloblo21}
   \end{center}
\end{figure}

\bigskip

\noindent{\bf LVE for Tensors.}
In the case of tensor models with quartic melonic interactions, 
the LVE intermediate fields are $D$ different collections of replica \emph{matrices} $\sigma^1 , \cdots \sigma^D$, 
each one acting on a particular colored strand of the ordinary tensor propagator.
For tensors one needs to improve the LVE \cite{Gurau} by adding an additional parametric representation for the 
corner resolvents 
\be   R   = \Big(1 + i H  \Big)^{-1}= \int_0^\infty e^{-\alpha  (1+iH )}  d \alpha \; .
\ee
This representation preserves the LVE convergence bounds, since $\Vert e^{-\alpha  (1+iH )}  \Vert \le e^{-\alpha }  $. But it 
has one additional remarkable property: it factorizes the action of the different colored intermediate matrix fields along 
\emph{well-defined colored faces}, see \eqref{traceform} below. These faces are in one to one correspondence 
with the connected components of the tree made of edges of a single color. In particular this allows one to recover the
correct scaling in $N$ of the tree amplitudes.

In order to explore subleading orders in $1/N$ one must use the mixed representation \cite{Gurau}. 
This consists in further Taylor expanding up to fixed order $2q$ in the coupling constant the product of resolvents.
 This further expansion rewrites any connected function of the theory 
in terms of trees decorated by up to $q$ loop edges, plus an integral remainder\footnote{This mixed expansion is closely related to the cleaning expansion
introduced in \cite{RZTW2,ZTW} to renormalize QFT in the LVE representation. It is not identical, as the cleaning expansion is organized along 
the Dyck path of the tree, whereas the mixed representation is not.}.

The main advantage of the tensor loop vertex representation and of its mixed 
expansion as defined in \cite{Gurau} is that now the leading order (indexed by trees) 
is exactly the leading term of the $1/N$ expansion, and the loop corrections with up to $q$ loop edges correspond to subleading 
terms up to order at least $N^{-q(D-2)}$ in the $1/N$ expansion. Hence the LVE is directly suited to the study of the $1/N$ expansion for 
tensors. Nevertheless not all loop corrections lead to the same combinatoric weights, or the same order in $1/N$. 
They hide a much richer variety of multiple scalings and critical points than matrix models. 
This paper is devoted to define and analyze the first infinite class of such subleading terms.

The full formalism for computing any cumulant of $\mu$ in \eqref{eq:measure}, i.e. all the Schwinger functions of the 
corresponding interacting random tensor theory has been developed in detail in \cite{Gurau}. It involves 
plane trees with ciliated vertices. The ciliated vertices  correspond to external faces, and in 
the case of many external legs the formalism requires some heavy
notations. For pedagogical reasons in this paper we shall treat only the two point function $\cG_2$ of the model, which
writes as a sum over plane trees $ \cT$ with a \emph{single ciliated vertex}. The extension of our results 
to the free energy and correlation functions of arbitrary order is direct and left as an exercise.   

The LVE expansion of the two point function is
\bea \label{eq:LVE}  
&& \cG_2 =  \sum_{n\ge 1} \frac{(-\lambda)^{n-1}}{n!} 
  \sum_{i_1, \cT} {\cA}(\cT) \, ,
\eea
where $i_1$ runs over the labels $1$ to $n$ and $\cT$ runs over the plane trees with $n$ vertices 
and oriented edges having a cilium on the vertex $i_1$. The contribution of a tree on $\cT$ is 
\bea \label{traceform}
&&  {\cA}(\cT ) =  \frac{1}{  N^{1+ n (D-1)} } \int_{0}^1 \Bigl( \prod_{ (i,j) \in \cT } du_{ij} \Bigr) \int d\mu_{w_{ij}^\cT (u)  1^{ \otimes D}  } (\sigma ) 
\crcr
&& 
\int \Big( \prod_{q=1}^{2n-2+1} d\alpha_q\Big)
\; e^{-\sum_{q=1}^{2n-2+1} \alpha_q } \prod_{f \in \cF( \cT ) }
\Tr \Big[ \prod^{\rightarrow}_{ q \in r(f^c)} e^{ -\alpha_q \sqrt{\frac{\lambda}{N^{D-1}}} (\sigma^{i(q)_c} - \sigma^{i(q)_c\dagger}) } \Big] .
\eea 
In this formula the parameters $u$ are associated to the $n-1$ edges of $\cT$.
The parameters $q$ runs over the $2(n-1)+1$ corners of the tree. In the tree $\cT$ there are $D$ single-colored
sub forests $\cF_c$. The sub forest $\cF_c ( \cT )$ is made of all edges in $\cT$ of color $c$, and the Dyck paths $r(f_c)$
around each of its connected components (including the connected components made of a unique vertex) 
form exactly the set of faces of color $i$. The full set of faces 
$\cF( \cT )$ is the union $\bigcup_{c=1}^D  \cF_c( \cT )$.

Although the intermediate matrix fields of different colors act on different strands they remain 
coupled in a non-trivial way since they share common $\alpha_q$ parameters (one per corner) and 
the non trivial Gaussian measure. This means that the quartic tensor model is not just a 
product of $D$ independent matrix models.

The $D$ collections of intermediate matrix fields $\sigma$
are distributed according to the covariances
\bea
&& \int d \mu_{w_{ij}^\cT(u)  1^{\otimes D} } (\sigma ) \; \; \sigma^{(k)_c}_{ab}  (\sigma^{(l)_{c'} \dagger})_{b'a'} =
w_{kl}^\cT (u ) \;  \delta_{aa'} \delta_{bb'} \delta^{cc'}\; .
\eea

The mixed expansion is written in terms of pairs $(\cT,\cal L)$ with $\cT$ a plane tree rooted at the cilium
and $\cal L$ a set of loop edges. A rooted plane tree decorated by loop edges is a map (i.e. a graph with an ordering of 
the half-edges at each vertex) with colored edges, which we denote $G =\pi_0 (\cT,\cal L) $. A map $G$ has $c$ sets of (possibly disconnected) 
sub maps $G_c$ made of the edges of color $c$ in $G$. $G_c$ also includes maps made of a unique vertex with no edges.
We denote the set of faces of each map $G_c$ by $\cF(G_c)$, and the set of faces of the map $G$ by 
$\cF(G) = \bigcup_{c}\cF(G_c)  \equiv \cF( \cT ,{\cal L}  )$.

\begin{theorem}[The Mixed Expansion\footnote{This expansion is called the mixed expansion because it is at the same time an 
expansion in $\lambda$ {\it and} an expansion in $1/N$. 
More precisely, being an expansion in $\frac{\lambda}{N^{D-2}}$, one can use it to establish the 
Borel summability of the free energy, or alternatively, one can use it to establish its $\frac{1}{N}$ expansion at all orders. }, \cite{Gurau}]\label{thm:mixedexp}
 The contribution of a tree ${\cA}(\cT ) $ admits an expansion in terms of trees 
decorated with a finite number of loop edges $ ( \cT, {\cal L}) $  
plus a Taylor remainder
\bea \label{treetaylor}
&&   {\cA}(\cT) 
   = \sum_{L=0}^{p-1} \cA^{(L)}   (  \cT   )
   + \cR^{(p)}  (  \cT)\;, \crcr
  &&  \cA^{(L)}  (  \cT   ) = 
\sum_{ {\cal L}, |{\cal L}| = L} \; \sum_{c_1\dots c_L }  \cA^{ (L) }( \cT,{\cal L}  )  
\eea  
where ${\cal L}$ runs over all possible ways to decorate $ \cT $ with $L$ oriented loop edges 
$l_1=(j_1,j_{1}')$, $l_2=(j_2,j_{2}')$ up to $l_L = (j_L,j_{L}')$ and $c_1,\dots c_L$ run over the possible colorings of the loop edges. 
The amplitude of a tree $\cT$ with $L$ loop edges forming the set $\cL$ is
\bea\label{eq:treecontrib}
 \cA^{(L) }   ( \cT, {\cal L}  )  =  N^{-1- (n+L)(D-1) +  | \cF ( \cT ,{\cal L}  ) |  }
 \frac{ (- \lambda )^L }{L!}
 \; \int_{0}^1   \Bigl( \prod_{ (i,j) \in \cT } du_{ij} \Bigr) 
 \prod_{L=1}^s w_{j_L  j_{L}'}^\cT (u)    \; .
\eea 
Remark that the power of $N$ exactly vanishes for melonic graphs which have $L=0$ and $| \cF ( \cT ,\emptyset ) | = D+ (D-1)(n-1) $.
The rest term is 
\bea 
 \cR^{(p)} (  \cT ) 
&=& \int_{0}^1 dt (1-t)^{p-1} \Bigg[ 
N^{ - 1 -(n+p)(D-1) }   \frac{(- \lambda )^p}{(p-1)!}   \sum_{ {\cal L}, | {\cal L} | = p } \;\;\sum_{c_1\dots c_p }  
 \int_{0}^1 \Bigl( \prod_{ (i,j) \in \cT } du_{ij} \Bigr) 
 \nonumber \\
&\times&  \prod_{s=1}^p w_{j_s j_s'}^\cT (u)  \int d\mu_{w_{ij}^\cT (u)  1^{ \otimes D}  } (\sigma )  
\int_0^{\infty} \Big( \prod_{q=1}^{2n-2+2p+1} d\alpha_q   \Big)  \; e^{-\sum_{q=1}^{2n-2+2p+1} \alpha_q }  
\nonumber\\
 &\times&   \prod_{f^c \in \cF ( \cT,{\cal L}  ) } \Tr \Big[ \prod^{\rightarrow}_{ q \in q(f^c)} 
 e^{ -\alpha_q \sqrt{t}\sqrt{\frac{|\lambda|}{N^{D-1}}} (\sigma^{i(q)_c} - \sigma^{i(q)_c\dagger}) } \Big] 
 \Bigg] \; . 
\eea 

The term $\sum_{L=0}^{p-1} \cA^{(L)}   (  \cT   )$ is a polynomial hence obviously analytic in $\lambda$. 
Furthermore the rest term in the mixed expansion is a convergent series in $\cT$,
hence defines a function analytic in the heart-shaped domain
$\lambda = e^{i \phi } \vert \lambda \vert$, $\vert \lambda \vert \le const.  \cos ^2\frac{\varphi}{2} $.
In this domain the following bounds hold \cite{Gurau}:
\bea
&& \big{|} \cA^{(q)}( \cT) \big{|} \le K \frac{|\lambda|^q}{N^{q(D-2)}} 
 \frac{ (2n+2q-2)! }{q! (2n-2)!} \\
&&  |\cR^{(p)} ( \cT ) |\le K \frac{1}{\bigl( \cos \frac{\varphi}{2} \bigr)^{2n+2p-1}  }
 \frac{|\lambda|^p}{N^{p(D-2)}} 
 \frac{ (2n+2p-2)! }{(p-1)! (2n-2)!} , \nonumber 
\eea 
with $K$ some constant depending only on $D$.
\end{theorem}

The map $G=\pi_0 (\cT,\cL)$ has at most $D+ (D-1)(n-1)+L$ faces. This explains the bounds on the remainder
and ensures that all the terms up to order $N^{-p(D-2)}$ are captured by summing up the contributions 
$ \cA^{(L) }   ( \cT, {\cal L}  )  $ with at most $p$ loops
\bea\label{eq:sumtreesloops}
\cG_2 && =  \sum_{n\ge 1} \frac{(-\lambda)^{n-1}}{n!} 
  \sum_{i_1, \cT} \sum_{L=0}^p\sum_{ {\cal L}, |{\cal L}| = L} \; \sum_{c_1\dots c_L }   N^{-1- (n+L)(D-1) +  | \cF ( \cT ,{\cal L}  ) |  } \crcr
  && \times \;
 \frac{ (- \lambda )^L }{L!}
 \; \int_{0}^1   \Bigl( \prod_{ (i,j) \in \cT } du_{ij} \Bigr) 
 \prod_{L=1}^s w_{j_L  j_{L}'}^\cT (u) \;\; +\, O(N^{-p(D-2)}) \;.
\eea 
Note that a map $ G =\pi_0 (\cT,\cL) $ might turn up to have less than $D+ (D-1)(n-1)+L$ faces, in which case it contributes
to an order lower than initially expected. The sum in  \eqref{eq:sumtreesloops} can be reorganized in terms of maps. Indeed, 
the contributions of couples  $(\cT,\cal L)$ coming from the same $\cT$ but such that the sets $\cL$ differ just by a permutation of 
the edges are equal and add up to cancel the $1/L!$ in \eqref{eq:sumtreesloops}. Furthermore, the same map $G$ is obtained by
completing any of its spanning trees by an appropriate set of loop edges (i.e. $\pi_0$ is onto but not one to one). However,
the factors $w(G,\cT)$ for a fixed $G$ add up to $1$, hence the sum in  \eqref{eq:sumtreesloops}
can be rewritten in terms of maps $G(i_1,n,L)$ having $n$ vertices labelled $1,2,\dots n$, rooted at $i_1$ and having $n-1+L$ colored 
oriented edges
\bea
 \cG_2 &=&  \sum_{L=0}^p \sum_{n\ge 1} \frac{(-\lambda)^{n+L-1}}{n!} \sum_{i_1,G(i_1,n,L)} N^{-1- (n+L)(D-1) +  | \cF \bigl( G(i_1,n,L) \bigr) | } \crcr
      &&+   O(N^{-p(D-2)})\; .
\eea 
Finally, grouping together the maps which differ just by a relabeling of the vertices or the orientation of the edges
we get a sum over rooted maps with $n$ unlabeled vertices and $n-1+L$ colored unoriented edges, 
\bea\label{eq:summaps}
 \cG_2 & =&  \sum_{L=0}^p \sum_{n\ge 1}  (-2 \lambda)^{n+L-1} \sum_{ G {\rm\; with\;} n+L-1 {\rm \; edges\; and \;} n {\rm\; vertices} }  N^{-1- (n+L)(D-1) + F(G) } \crcr
      && +   O(N^{-p(D-2)})\; ,
\eea 
where $ F(G) =| \cF ( G )  |$.
This in particular explains why $z=-2\lambda$ is a good variable to consider in the sequel. The $-$ signs reflects that stability of the theory 
at $\lambda >0$; it also means that the 
singularity of $\cG^{0}_2 (\lambda)$ is at $\lambda = \lambda_c = -(8D)^{-1}$, hence on the
negative side of the real axis, which is not in the heart-shaped constructive domain of convergence.
This major problem is not tackled in this paper but kept for future study.

\section{Graph classification and the main result}

\label{sec:GraphandTheo}

\subsection{Pruning and Grafting}

We can work thus either at  the constructive level, which requires the LVE trees or at the perturbative with maps $G$. From now on 
we no longer care about the remainder term  $\cR$: we sent $p \to \infty$ in \eqref{eq:summaps} and ignore 
the fact that the perturbative sum over $L$ from $0$ to $\infty$ is not summable. 

From now on we denote $V(G)=n$ the number of vertices, $E(G)=n-1+L$ the number of edges and $F(G)$ the number of faces of the map $G$. The ciliated
vertex in $G$ corresponds to the external faces of the two point function $\cG_2$.

To every map $G$ we will associate a map $\bar G$ which captures all its essential characteristics. The map $\bar G$
is obtained through the operations of {\it pruning} and {\it reduction} defined below. There exists an infinity of maps $G$ which 
correspond to the same $\bar G$. All such $G$ sum together
and yield a contribution associated to $\bar G$. The perturbative series can then be re-indexed in terms
of $\bar G$ and the double scaling limit is analyzed in terms of $\bar G$. Our classification is 
closely related to Wright's approach \cite{wright1,wright2} to the enumeration 
of labelled graph with a fixed number of loop edges.

We now remove iteratively all non-ciliated vertices of coordination one of $G$. This is called {\it pruning}.
For simplicity, in the sequel we shall refer to maps as graphs.  

\begin{definition}
A \emph{reducible leaf} of $G$ is a vertex of $G$ of coordination 1 which is not the ciliated vertex.
The \emph{pruned graph} $\tilde G $ associated to $G$ is obtained by removing inductively all 
reducible leaves and their unique attaching edge.
\end{definition}

\begin{figure}[ht]
   \begin{center}
 \includegraphics[width=14cm]{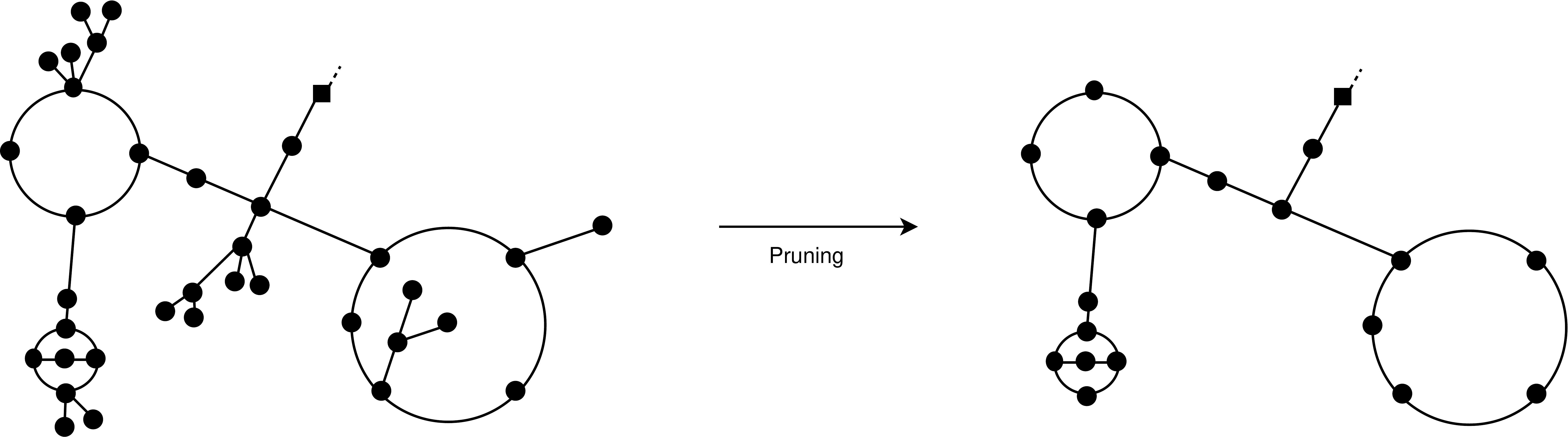}
 \caption{Pruning}
  \label{bloblo1}
   \end{center}
\end{figure}

The pruned graph is therefore obtained by removing from $G$ all rooted tree subgraphs with only reducible leaves,
see Fig. \ref{bloblo1}. Again the map $\pi_1 : G \to \pi_1 (G) = \tilde G$ is onto but not one to one. Remark that the pruned graph 
$\tilde G$ is never empty, as it contains at least the ciliated vertex. It also contains $L$ independent cycles, like $G$. 
Remark also that pruning is compatible with the coloring: the initial graph $G$ is colored, and the
corresponding pruned graph $\tilde G$, which is a subgraph of $G$, is also colored.

\begin{figure}[ht]
   \begin{center}
 \includegraphics[width=12cm]{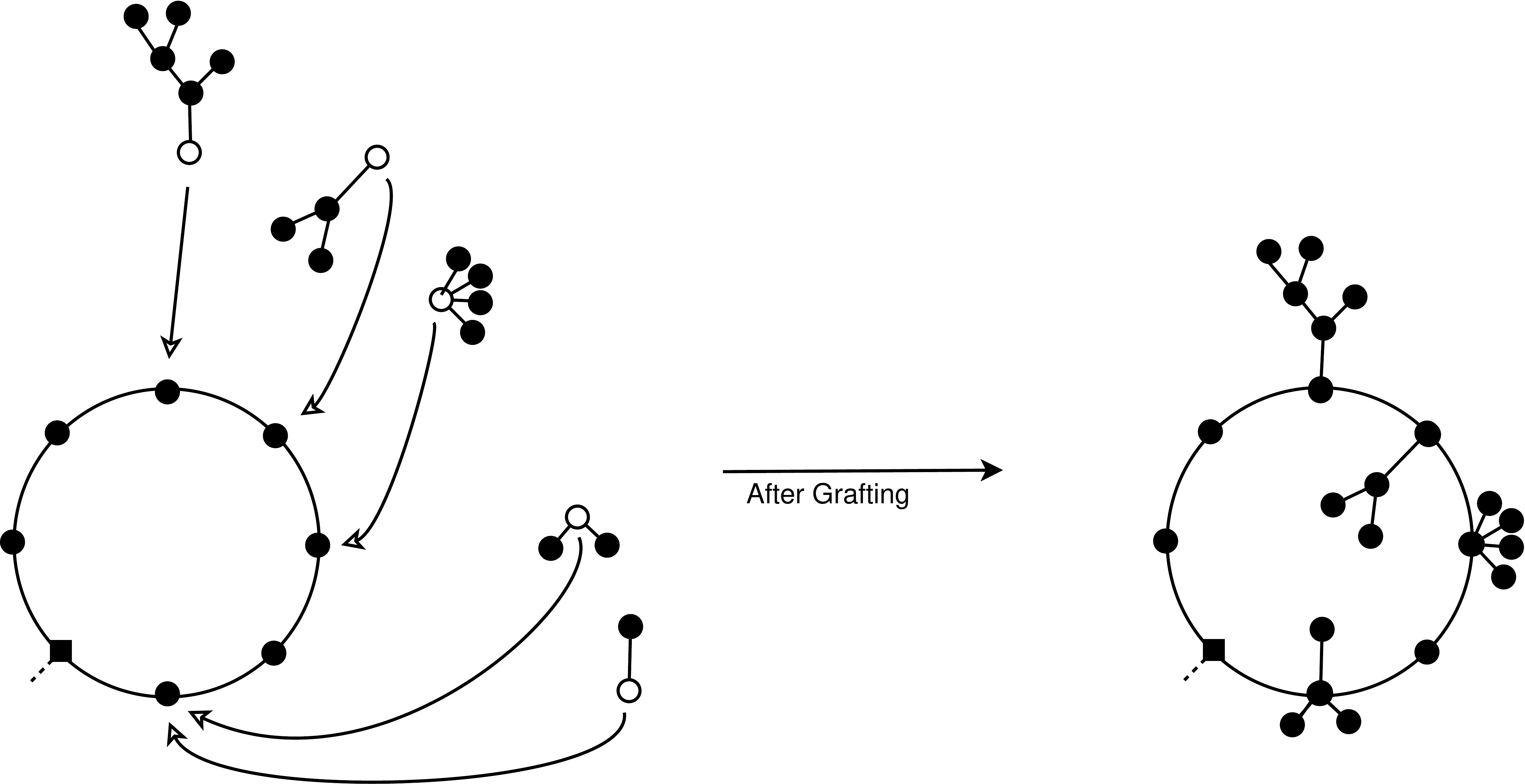}
 \caption{Grafting}
  \label{bloblo2}
   \end{center}
\end{figure}

The ``inverse'' operation of pruning is grafting: the initial graph $G$ is obtained by grafting some 
(possibly empty) rooted plane tree on each \emph{corner} of the pruned graph $\tilde G$ (see Fig. \ref{bloblo2}).
Considering all the possible graftings on $\tilde G$ reconstructs the entire family of graphs $G$ which reduces to $\tilde G$
by pruning.

Remark that the pruned graph $\tilde G$ has fewer vertices and fewer edges than $G$. However, at every step in the pruning
process the number of edges and vertices of the graph decreases by $1$, hence $\tilde G$ and $G$ have the same number of cycles
\bea
L(\tilde G)= E(\tilde G) - V(\tilde G) +1 = E(G) - V(G) +1=L(G) \; .
\eea 

Furthermore, all the graphs $G$ corresponding to the same pruned graph $\tilde G$ have the same scaling in $N$.
Indeed, when deleting an univalent vertex (and the edge connecting it to the rest of the graph, say of color $c$) 
the number of faces of the graph decreases by $D-1$, as all the faces of color $c\neq c'$ containing the vertex are deleted,
but the face of color $c$ is not. Hence
\bea
 -1- \bigl(E(G)+1 \bigr)(D-1) +   F ( G )  = -1  - \bigl( E(\tilde G)+1 \bigr)(D-1) +  F ( \tilde G )   .
\eea

\begin{definition} A \emph{one particle irreducible component} $\tilde G_k$ (1PI) of a pruned graph $\tilde G$ is a maximal (non empty) connected set of edges,
together with their attached vertices, which cannot be separated into two connected components by deleting one of its edges.
The ciliated vertex can either be part of a one particle irreducible component, or not, in which case it is called bare.
An \emph{irreducible component} of a pruned graph $\tilde G$ is either a one particle irreducible component $\tilde G_k$, or the 
ciliated vertex if it is bare.
\end{definition}

Remark that since each one particle irreducible component must include at least one loop edge (i.e. it must have at least a cycle), 
the number $p(\tilde G)$ of irreducible components in a pruned graph $\tilde G$
with $L$ loops is at least $1$ and at most $L+1$, and if $p(\tilde G) = L+1$, the ciliated vertex must be \emph{bare}.

\subsection{Reduction}

We now remove all non-ciliated vertices of coordination 2. This is called \emph{reduction}.

\begin{definition}
A vertex of a pruned graph is called \textsl{essential} if it is of degree strictly greater than 2 or if  it is the ciliated vertex.
A bar of a pruned graph is a maximal chain of edges with internal vertices all of degree 2. The number of essential vertices 
and of bars of a pruned graph $\tilde G$ will be noted $V^{e}(\tilde G)$ and $B(\tilde G)$.
\end{definition}

\begin{figure}[ht]
   \begin{center}
 \includegraphics[width=12cm]{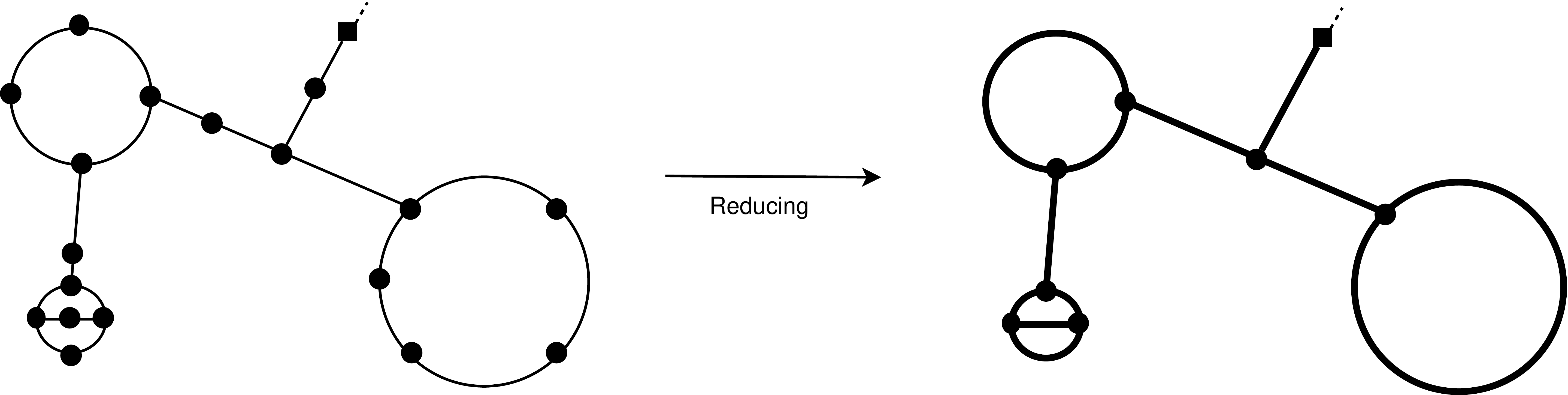}
 \caption{Reduction of a pruned graph}
  \label{bloblo3}
   \end{center}
\end{figure}

If we picture each bar of a pruned graph $\tilde G $ as a (fat) edge, we obtain a new graph $\bar G$ associated to $\tilde G$, which is made of 
$V(\bar G)=V^{e}(\tilde G)$ vertices plus $E(\bar G) = B(\tilde G)$ (fat) edges between them  (see Fig. \ref{bloblo3}). It has 
still the same number $L(\bar G) = L(\tilde G) = L(G)\equiv L$ of independent cycles as $\tilde G$ and $G$, 
and every tree of $\bar G$ has $E(\bar G) - L$ edges.  

\begin{lemma}\label{barbound}
We have $E(\bar G) \le 3L -1$, and $V(\bar G) \le 2L$. Moreover $E(\bar G) = 3L -1$ implies that every vertex of $\bar G$ is of degree 3, except the ciliated vertex 
$v_c$ which is of degree 1.
\end{lemma}
\prf  Let $d_v$ be the degree of vertex $v \in \bar G$. We have $d_v \ge 3$ for any $v$ except possibly the ciliated vertex, with degree $d_c$.
Furthermore $V(\bar G)-1 + L = E(\bar G)$, as $\bar G$ can always be decomposed into a tree of $V(\bar G)-1$ edges plus a set of $L$ loop edges.
We have 
\be 2E(\bar G) = \sum_v d_v \ge 3(V(\bar G)-1) +d_c = 3(E(\bar G)-L) +d_c
\ee
which proves that $E(\bar G) \le 3L -d_c \le 3L-1$. Equality can happen only if $d_v =3$ for $v \ne v_c$ and $d_c =1$. Finally
since $\bar G$ is connected $E(\bar G) = V(\bar G) -1 + L$ , hence $E(\bar G)\le 3L-1 $ implies $V(\bar G) \le 2L$. 
\qed

\medskip

If a chain of edges of $\tilde G$ which we reduce is made of edges all with the same color $c$, we color the new fat edge in $\bar G$ with $c$. If on the contrary 
the chain of edges contained at least two edges of two different colors, we will call the fat edge multicolored, and we will associate to it an index $m$. 

\begin{definition} A reduced graph $\bar G$ is a regular graph with one ciliated vertex, such that all other vertices
have coordination at least 3, plus the choice of a color $c$ or a label $m$  on any of its edges. Forgetting the labels, one gets 
an associated unlabeled reduced graph.
\end{definition}

The reduction operation $\pi_2$ is now well-defined from the category of (colored) pruned graphs to the category of reduced graphs.
The map $\pi_2 : \tilde G \to \pi_2 (\tilde G) = \bar G$ is onto but not one to one.
The reverse of the reduction map, called \emph{expansion}, expands any edge into an arbitrarily long chain with degree 2 intermediate vertices,
and sums over colorings compatible with the colors or the label $m$.

We denote $E^m(\bar G)$ the number of multicolored edges of the reduced graph $\bar G$. We define the (possibly disconnected)
graph $\bar G_c$ as the subgraph of $\bar G$ made of all the vertices of $\bar G$ and all the edges of color $c$.
As $\bar G_c$ is a map, it has a certain number of faces, $F(\bar G_c)$, and a total genus $g(\bar G_c)$ (i.e. the sum of the genera of
the connected components of $\bar G_c$). We denote the number of cycles of $\bar G_c$ by 
$L(\bar G_c)$.
\begin{lemma}\label{lem:Nscaling}
 The scaling with $N$ of all the pruned graphs associated to the reduced graph $\bar G$ is
 \bea
   -D L(\bar G) + 2 \sum_c L( \bar G_c) -2 \sum_c g( \bar G_c) \; .
 \eea 
\end{lemma}

\prf The scaling with $N$ of a pruned graph $\tilde \cG$ is 
\bea
-1  - \bigl( E(\tilde G)+1 \bigr)(D-1) +  F(\tilde G ) \;.
\eea 
When deleting a bivalent vertex on an unicolored edge of color $c$ the number of edges of the graph decreases by 1 and the number of faces 
by $D-1$ (one for each $c'\neq c$). On the contrary,  when deleting all the bivalent vertices on a multicolored edge with $p$ intermediate 
vertices, the number of edges decreases by $p$, and the number of faces by $D-2 + (D-1)(p-1)$, hence the scaling with $N$ writes in terms of the data
of the reduced graph as
\bea
 -1 - \bigl( E(\bar G)+1 \bigr)(D-1) - E^m(\bar G) + \sum_c F(G_c)\; .
\eea 
The graphs $G_c$ have $C(G_c)$ connected components, $V(G_c)$ vertices, $E(G_c)$ edges and total genus $g(G_c)$, hence the scaling with $N$ writes
\bea
 && -1 - \bigl( E(\bar G)+1 \bigr)(D-1) - E^m(\bar G) \crcr
 && + \sum_c \bigl( -V(G_c) + E(G_c) + 2C(G_c) - 2g(G_c)\bigr) \; .
\eea 
The number of connected components of $G_c$ can be computed in terms of the number of cycles, $E(G_c) = V(G_c) - C(G_c) + L(G_c)$,
and we get 
\bea
&& -1 - \bigl( E(\bar G)+1 \bigr)(D-1) - E^m(\bar G) \crcr
 && \quad  +\sum_c \bigl[ -V(G_c) + E(G_c) + 2 \bigl(V(G_c) - E(G_c) + L(G_c)\bigr) - 2g(G_c)\bigr] \crcr
&&=  -1 - \bigl( E(\bar G)+1 \bigr)(D-1) - E^m(\bar G) \crcr
&& \quad +\sum_c \bigl( V(G_c)  - E(G_c) + 2 L(G_c) - 2g(G_c)\bigr) \; .
\eea 
As $V(G_c) = V(\bar G)$ and $E(\bar G) = E^m(\bar G) + \sum_c E(G_c)$ we rewrite this as
\bea 
 -1 - \bigl( E(\bar G)+1 \bigr)(D-1) - E(\bar G) + D V(\bar G) + \sum_c \bigl( 2 L(G_c) - 2g(G_c)\bigr) \;, 
\eea 
and using  $E( \bar G ) = V( \bar G ) -1 + L(\bar G)$ the lemma follows.

\qed

Summing the family of graphs which, through pruning and reduction, lead to the same reduced graph $\bar G$ we obtain the {\it amplitude} of $\bar G$.
\begin{theorem}\label{thm:ampli}
 The amplitude of a reduced graph $\bar G$ is 
 \bea \label{eq:ampli}
  \cA(\bar G) &=& N^{-D L(\bar G) + 2 \sum_c L( \bar G_c) -2 \sum_c g( \bar G_c)} \; T(Dz)^{1+2E(\bar G)}\\
  &&  \Bigl( \frac{ z }{1-zT^2(Dz)} \Bigr)^{\sum_{c } E( \bar G_c) }
  \Bigl(  \frac{ D(D-1) z^2 T^2(Dz)}{[ 1-DzT^2(Dz) ][ 1-zT^2(Dz)  ] } \Bigr)^{E^{\text{m}}(\bar G)} \; .\nonumber
 \eea 
\end{theorem}
 
 The proof of this theorem is presented in section \ref{sec:examples}.

A further simplification comes from the fact that the 1PR bars of $\bar G$ do not in fact need any label $c$ or $m$.
This comes from the fact that such graphs have the same scaling in $N$, but 
the graphs with colored $1PR$ edges are suppressed in the double scaling limit: one can  chose to group together 
or not the graphs which differ only by the label of their 1PR bars without changing the double scaling limit. 
Indeed, if one combines together all the reduced graphs differing only by the labels of the 1PR edges (which in 
this case we call free) one gets a contribution
 \bea
 &&N^{-D L(\bar G) + 2 \sum_c L( \bar G_c) -2 \sum_c g( \bar G_c)} \; T(Dz)^{1+2E(\bar G)}\crcr
  &&  \Bigl( \frac{ z }{1-zT^2(Dz)} \Bigr)^{\sum_{c } E( \bar G_c) }
  \Bigl(  \frac{ D(D-1) z^2 T^2(Dz)}{[ 1-DzT^2(Dz) ][ 1-zT^2(Dz)  ] } \Bigr)^{E^{\text{m} }  (\bar G)   } \crcr
  &&   \Bigl(  \frac{ D  z }{[ 1-DzT^2(Dz) ]  } \Bigr)^{E^{\text{free}}(\bar G)} \; ,
 \eea 
having the same critical behavior as  \eqref{eq:ampli}, where $ E^{\text{free }}(\bar G) $  denotes the number of free edges of $\bar G$
and $E^{\text{m} }  (\bar G)$ the number of multicolored edges in $\bar G$.  
 
\subsection{Cherry Trees}

\begin{definition} \label{defcherry}   
A reduced graph with $L(\bar G)$ loops and one cilium is called a \emph{cherry tree} (or, in short a \emph{cherry}) if 
$E(\bar G) = 3L(\bar G)-1$, $L(\bar G) = \sum_c L(\bar G_c)$, and $V(G_c) = C(G_c)$. 
\end{definition}

\begin{figure}[ht]
   \begin{center}
 \includegraphics[width=4cm]{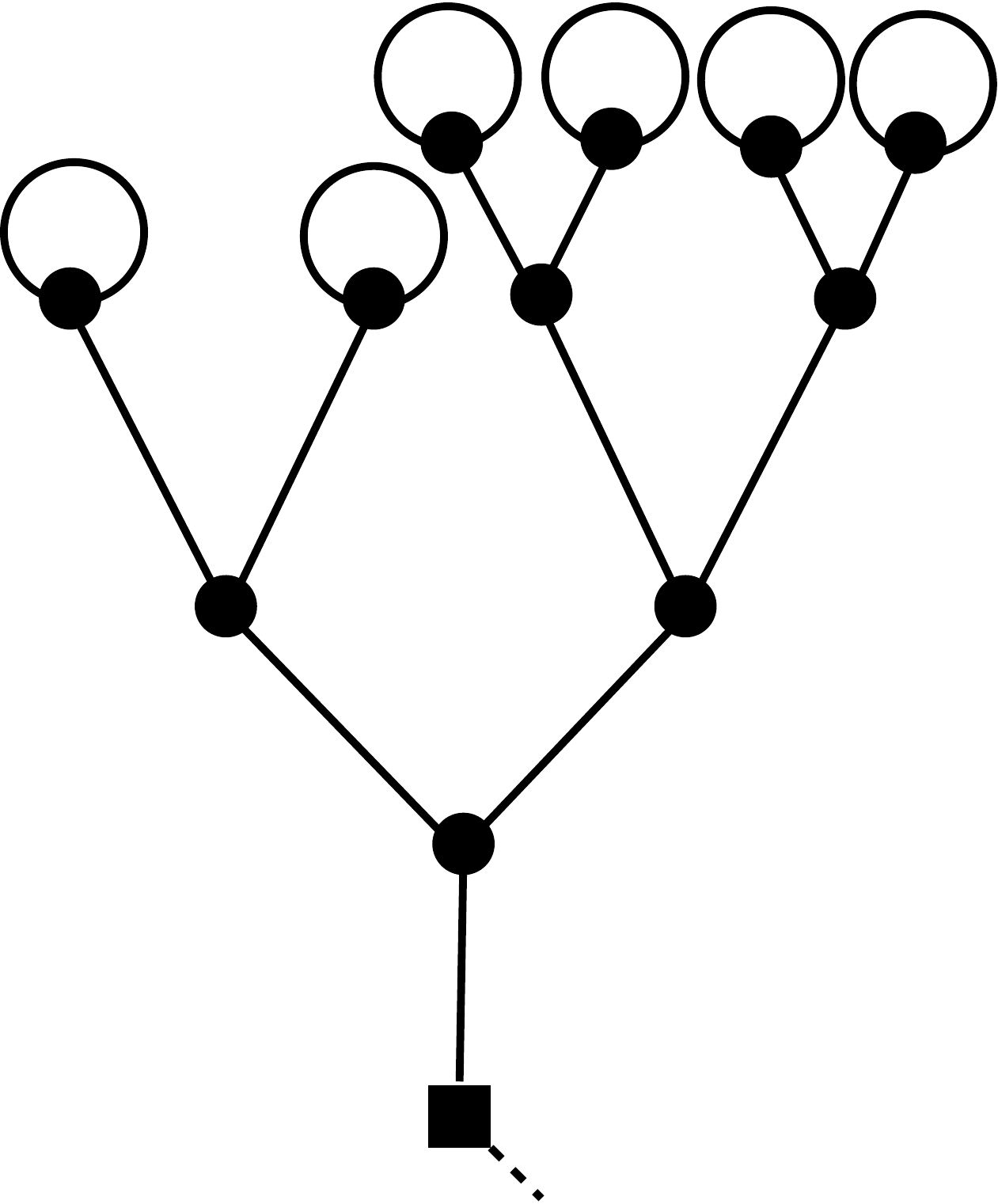}
 \caption{An example of cherry tree}
  \label{bloblo4}
   \end{center}
\end{figure}

\begin{figure}[ht]
   \begin{center}
 \includegraphics[width=8cm]{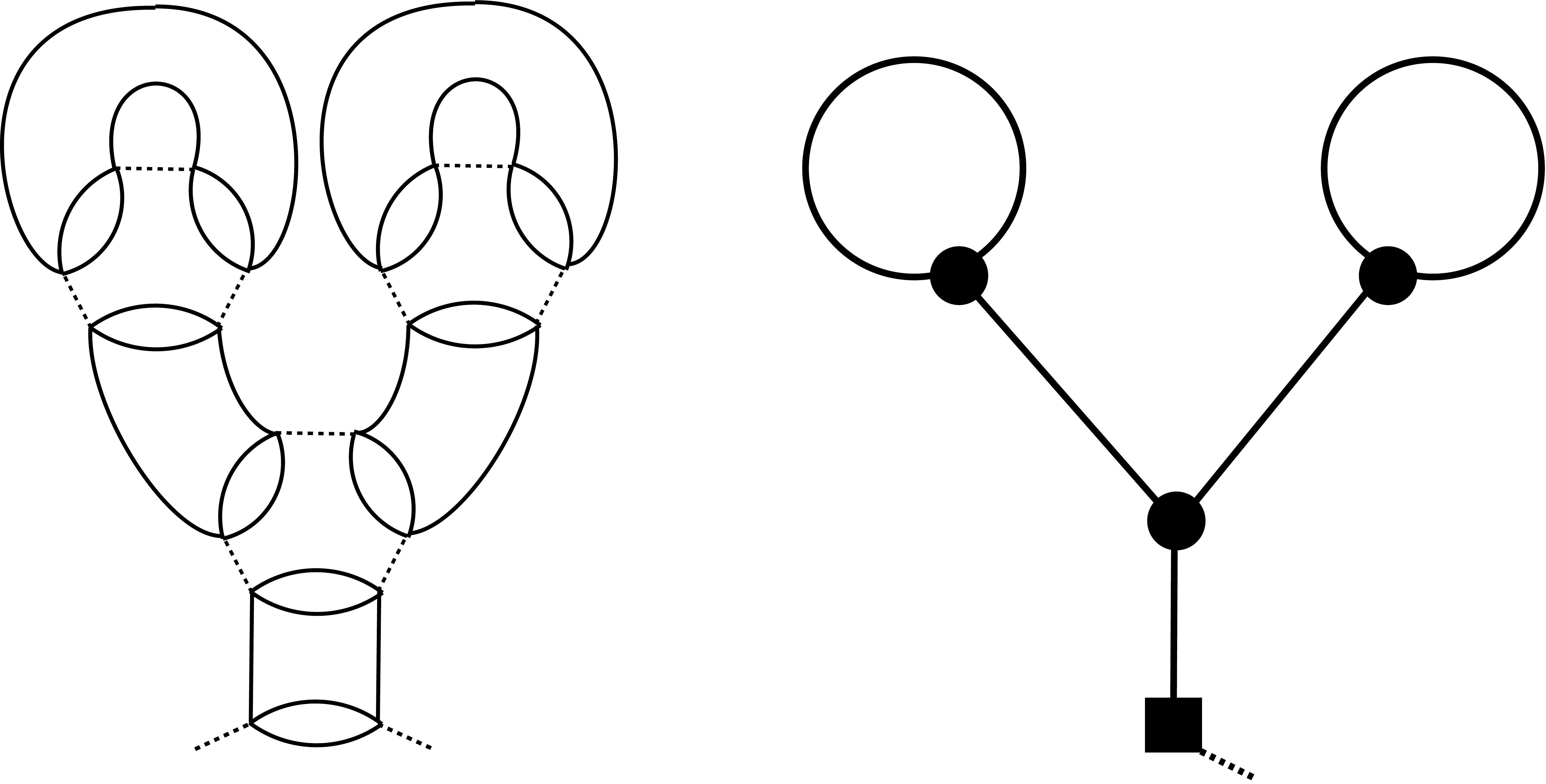}
 \caption{A cherry tree in the direct and LVE representations}
  \label{blobloblo}
   \end{center}
\end{figure}

Remark that such a graph is a rooted binary tree made of multicolored edges, with $L(\bar G)+1$ univalent vertices (the leaves and the root) and $L(\bar G)-1$
trivalent vertices decorated by one self loop of color $c=1,\dots D$ on each of its leaves. Indeed, $E(\bar G) = 3L(\bar G)-1 $
implies that the root vertex is univalent and all other vertices are trivalent, $V(G_c) = C(G_c) $ implies that all the connected components
of $G_c$ have exactly one vertex, $L(\bar G) = \sum_c L(\bar G_c) $ and the fact that all vertices except the root are trivalent 
implies on one hand that $L(\bar G)  $ vertices are decorated by self loops of a fixed color, and on the other that the graph obtained by erasing these
self loops has no more cycles and is connected, hence is a tree (see Fig. \ref{bloblo4} and   \ref{blobloblo}).

The mixed pruned expansion for the two-point function then splits into two parts, the sum over cherry trees
and the rest, which is the sum over all other graphs:
\be  \cG_2^{L} (z,N)=  \cG_{2,cherry}^{L}(z,N) + \cG^{L}_{2,rest}(z,N) .
\ee
The next step is to change $z$ to the \emph{rescaled variable} $x = N^{D-2}(z_c-z)$ with $z_c = (4D)^{-1}$,
\be  \cG_{2,cherry}^{L}(z,N) =   \cG_{2,cherry}^{L,x}(N) ; \;  \cG^{L}_{2,rest}(z,N) = \cG^{L,x}_{2,rest}(N) 
\ee
and to consider now at \emph{fixed $L$ and $x$} the asymptotic expansion of these quantities when $N \to \infty$.

Our main result is that in this regime the cherry trees contribution $ \cG_{2,cherry}^{L,x}(N)$ has a leading
term proportional to $N^{1-D/2}$ with a coefficient which can be computed \emph{exactly}, and that 
for $3 \le D \le 5$ this leading term dominates all the rest by at least one additional $N^{-1/2}$ factor:

\begin{theorem}[Main Bound] 
\label{maintheo} 
For $3\le D \le 5 $, $L \ge 1$ fixed and for $x$ in some neighborhood of $x_c = \frac {1}{4(D-1)}$, we have 
\be
 \label{mainequa}  \cG_{2,cherry}^{L,x} (N)  =  N^{1-D/2} \frac{8\sqrt D  C_{L-1}}{ [16(D-1)]^L}  \;  x^{-L +1/2} + O(  N^{1/2-D/2} ) \; ,
\ee 
where $C_{L-1} = \frac{1}{2L-1} \binom{2L-1}{L-1} $ is the Catalan number of order $L-1$. Moreover
there exists a constant $K_L$  such that 
\be \vert  \cG^{L,x}_{2,rest}(N) \vert \le  N^{1/2-D/2}  \;    K_L \;  x^{-\frac{3}{2}L+\frac{1}{2}} \label{maininequa}   \; .
\ee
\end{theorem}

Thus for $D <6$ the cherry trees yield the leading contribution in the double scaling limit $N\to \infty$, $z\to z_c$, $x = N^{D-2}(z_c-z)$ fixed.
However remark that there \emph{is a certain degree of arbitrariness} in the separation between cherry trees and the rest.
The cherry trees are just the simplest convenient sub series displaying the leading double scaling behavior.
The rest term is less singular, which means that adding part of it to the cherry trees would typically not change the nature of the singularity.
However adding \emph{all} graphs at once would certainly at some point meet the problem of divergence of perturbation theory\footnote{Hence analytic 
continuation of the Borel sum might be the ultimate justification of (multiple) scalings.}.

\section{Reduced graphs amplitudes}\label{sec:examples}

We prove in this section Theorem \ref{thm:ampli}.

The family of pruned graphs associated to $\bar G$ is obtained by arbitrary insertions of bivalent vertices on any of its edges. The grafting operation adds 
arbitrary trees with colored edges independently on each of the corners of the pruned graphs. From \eqref{eq:summaps} each edge of a graph has a weight $z$.

We denote $T(z)$ the generating function of plane trees with weight $z$ per edge
\bea
 T(z) = \sum_{n\ge 0} \frac{1}{2n+1} \binom{2n+1}{n}  z^n \; ,\quad T(z) = 1 + zT^2(z)   \; ,\quad  T(z)  = \frac{1 - \sqrt{1-4z}}{2z}  \; .
\eea 
As the edges of the LVE trees have a color $1,\dots D$ on each edge, their generating function is $T(Dz)$. We have 
\bea\label{eq:smecher}
 T(Dz) = 1 + Dz T^2(Dz) \Rightarrow 1-DzT^2(Dz)=2-T(Dz)=T(Dz) \sqrt{1-4Dz} \; .
\eea 

Every corner of a pruned graphs represents a place where a tree can be inserted, hence has a weight $T(Dz)$. 
Exactly $1+2E(\bar G)$ of the corners of the pruned graph $\tilde G$ appear in the reduced graph $\bar G$, hence the corners 
of $\bar G$ bring in total a factor
\bea\label{eq:corners}
 T(Dz)^{ 1+2E(\bar G)} \; .
\eea 

The weight of the edges of the reduced graph are obtained by summing over the number of intermediate bivalent vertices in the associated pruned graphs 
from $0$ to infinity. Every intermediate vertex brings two new corners and a new edge, hence a factor $zT(Dz)^2$. 
As a function of the label $c$ or $m$ of the edge in $\bar G$ we distinguish several cases:
\begin{itemize}
 \item The edge in $\bar G$ has a color $c$. Then all the intermediate edges in $\tilde G$ have the same color, and we get a total weight
  \bea\label{eq:cedges}
    \sum_{k=0}^{\infty} z^{k+1} T^{2k}(Dz) =  \frac{ z }{1-zT^2(Dz)} 
  \eea 
  \item The edge in $\bar G$ has an index $m$. Then there must be at least two edges of different colors in $\tilde G$ which are replaced by the edge $m$, hence 
  the weight is
  \bea\label{eq:medges}
    && \sum_{k=0}^{\infty} D^{k+1} z^{k+1} T^{2k}(Dz) - D  \sum_{k=0}^{\infty} z^{k+1} T^{2k}(Dz) \crcr
    && = \frac{ D z }{1-DzT^2(Dz)} -  \frac{ D z }{1-zT^2(Dz)} \crcr
    && = \frac{ D(D-1) z^2 T^2(Dz)}{[ 1-DzT^2(Dz) ][ 1-zT^2(Dz)  ] } \; .
  \eea 
\end{itemize}
Note that if we were to consider the 1PR edges free, there weight becomes
\bea
  \frac{ D z }{1-DzT^2(Dz)} \; .
\eea 
Putting together the scaling in Lemma \ref{lem:Nscaling} with the contributions of  \eqref{eq:corners}, \eqref{eq:cedges} and \eqref{eq:medges}
proves Theorem \ref{thm:ampli}. We present here some low order examples of reduced graphs and their amplitudes.

\bigskip

\noindent{\bf Zero Loop.} At the leading order in $1/N$ only the trees with zero loops contribute. By pruning  
they reduce to the single ciliated vertex, which we call the graph $\bar G_0$\footnote{The reduction is trivial in this case 
as there are no two valent vertices in the pruned graph.}. 
The series of graphs $G$ corresponding to this trivial pruned graph is obtained by grafting a plane tree with colored edges 
on the single corner of the ciliated vertex, hence
\bea
 \cA(\bar G_0)= T(Dz) \; .
\eea 
reproducing the result of \cite{Gurau}.

\bigskip

\begin{figure}[ht]
   \begin{center}
 \includegraphics[width=4cm]{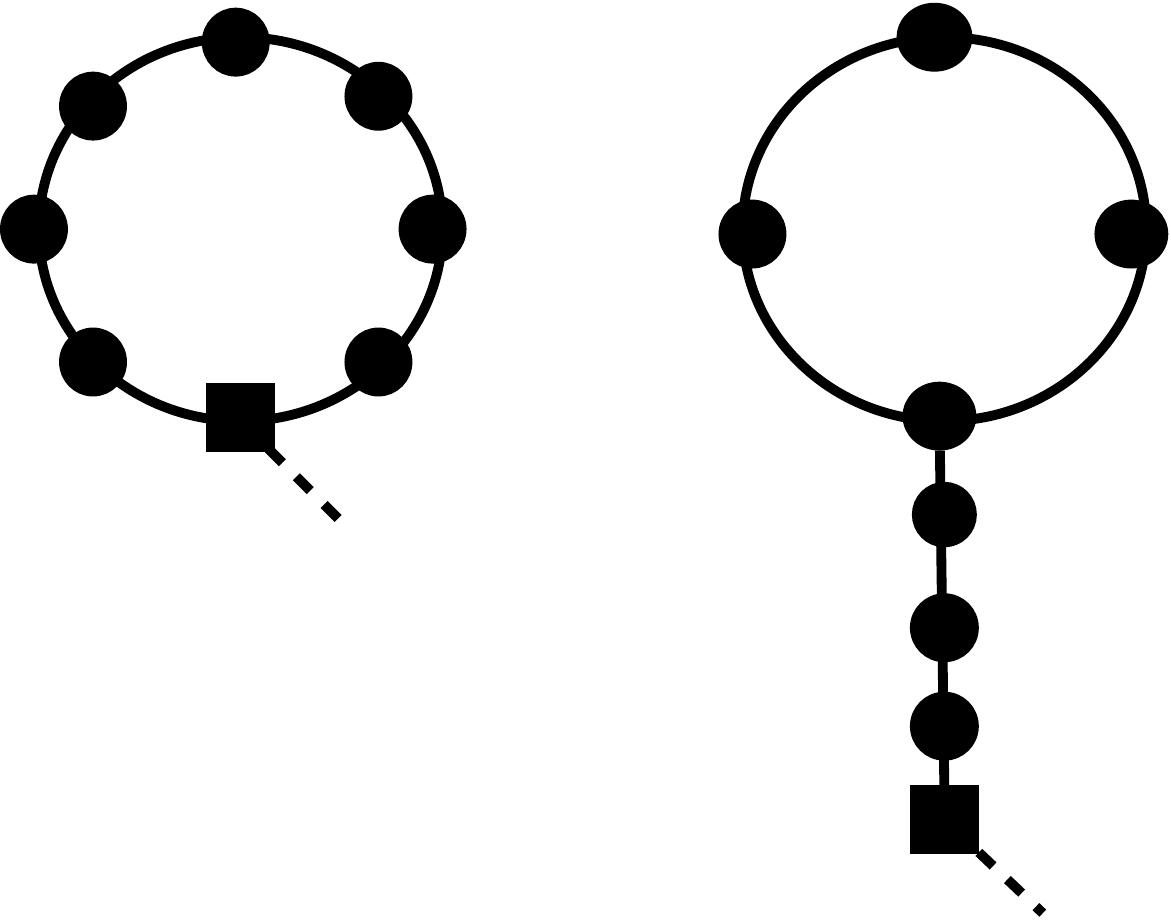}
 \caption{Two pruned graphs in the 1 loop case}
  \label{1looppruned}
   \end{center}
\end{figure}

\noindent{\bf One Loop.} The first $1/N$ corrections arise from graphs having one loop edge.  
There are two reduced graphs, see Fig. \ref{1looppruned} showing two of the pruned graphs corresponding to each of the two reduced graphs.

Consider the leftmost graph. If the loop edge in $\bar G$ has a color $c$ we get an amplitude
\bea
 \frac{1}{N^{D-2}} T(Dz)^3 \frac{ z }{1-zT^2(Dz)} \; ,
\eea 
while if the loop edge is multicolored we get an amplitude
\bea
 \frac{1}{N^D} T(Dz)^3 \frac{ D(D-1) z^2 T^2(Dz)}{[ 1-DzT^2(Dz) ][ 1-zT^2(Dz)  ] } 
\eea 

For the graph on the right hand side the loop edge and the 1PR edge can be either unicolored or multicolored. Denoting $l$ the self loop and
$t$ the vertical edge we get in each case the amplitude
\bea
&& l = c; t=c \qquad \frac{1}{N^{D-2}} T(Dz)^{5} \Bigl(  \frac{ z }{1-zT^2(Dz)}  \Bigr)^2 \crcr 
&& l =c; t =c'\neq c \qquad \frac{1}{N^{D-2}} T(Dz)^{5} \Bigl(  \frac{ z }{1-zT^2(Dz)}  \Bigr)^2 \crcr 
&&  l =c; t=m \qquad \frac{1}{N^{D-2}} T(Dz)^{5} \Bigl(  \frac{ z }{1-zT^2(Dz)}  \Bigr) \Bigl( \frac{ D(D-1) z^2 T^2(Dz)}{[ 1-DzT^2(Dz) ][ 1-zT^2(Dz)  ] } \Bigr) \crcr 
&&   l =m; t=m \qquad \frac{1}{N^{D}} T(Dz)^{5} \Bigl( \frac{ D(D-1) z^2 T^2(Dz)}{[ 1-DzT^2(Dz) ][ 1-zT^2(Dz)  ] } \Bigr)^2  \; . 
 \eea 

The contribution of these graphs is obtained by summing over the choices of colors $c$ and $c'$
\bea
 &&  \frac{D}{N^{D-2}} T(Dz)^{5} \Bigl(  \frac{ z }{1-zT^2(Dz)}  \Bigr)^2 + \frac{D(D-1)}{N^{D-2}} T(Dz)^{5} \Bigl(  \frac{ z }{1-zT^2(Dz)}  \Bigr)^2 \crcr
                && + \frac{D}{N^{D-2}} T(Dz)^{5} \Bigl(  \frac{ z }{1-zT^2(Dz)}  \Bigr) \Bigl( \frac{ D(D-1) z^2 T^2(Dz)}{[ 1-DzT^2(Dz) ][ 1-zT^2(Dz)  ] } \Bigr) \crcr
                && + \frac{1}{N^{D}} T(Dz)^{5} \Bigl( \frac{ D(D-1) z^2 T^2(Dz)}{[ 1-DzT^2(Dz) ][ 1-zT^2(Dz)  ] } \Bigr)^2 .
\eea 
Taking into account that 
\bea 
1-DzT^2(Dz) =T(Dz) \sqrt{1-4Dz} \; , \qquad T(Dz)\to_{z\to (4D)^{-1} } 2 \; ,
\eea 
and summing up all contributions, the one loop correction to the two point function exhibits the critical behavior
\bea
 \cG^1_{2}(z)  \sim \frac{1}{N^{D-2}} \frac{D}{(D-1)   \sqrt{1-4Dz}  }  +\frac{1}{N^D}  \frac{1}{ 2 (1-4Dz) } .
\eea 
The first term reproduces the results found in \cite{KOR}: the non analytic behaviour of the first correction in $1/N$ has a
susceptibility exponent $ \cG^1_{2}(z)  \sim (z_c-z)^{1-\gamma}  $, $\gamma = +3/2$. 
This can be rewritten as 
\be
\cG^1_{2}(z) \sim \frac{1}{N^{\frac{D-2}{2}}} \Bigl( \frac{\sqrt{D} }{2(D-1) \sqrt{ N^{D-2} [(4D)^{-1}-z] } } 
+ \frac{ N^{\frac{D-6}{2}} }{8D  N^{D-2}[(4D)^{-1}-z]  } \Bigr) \; ,
\ee
hence the second term is washed out for $D<6$ in the double scaling limit 
$N\to \infty$, $z\to (4D)^{-1}$, $ N^{D-2} [(4D)^{-1}-z]  $ fixed.

\bigskip

\noindent{\bf Two Loops.} There are several graphs contributing at two loops whose edges can furthermore be unicolored or multicolored. 
We will analyze here only some relevant examples.
\begin{figure}[ht]
\begin{center}
\includegraphics[width=7cm]{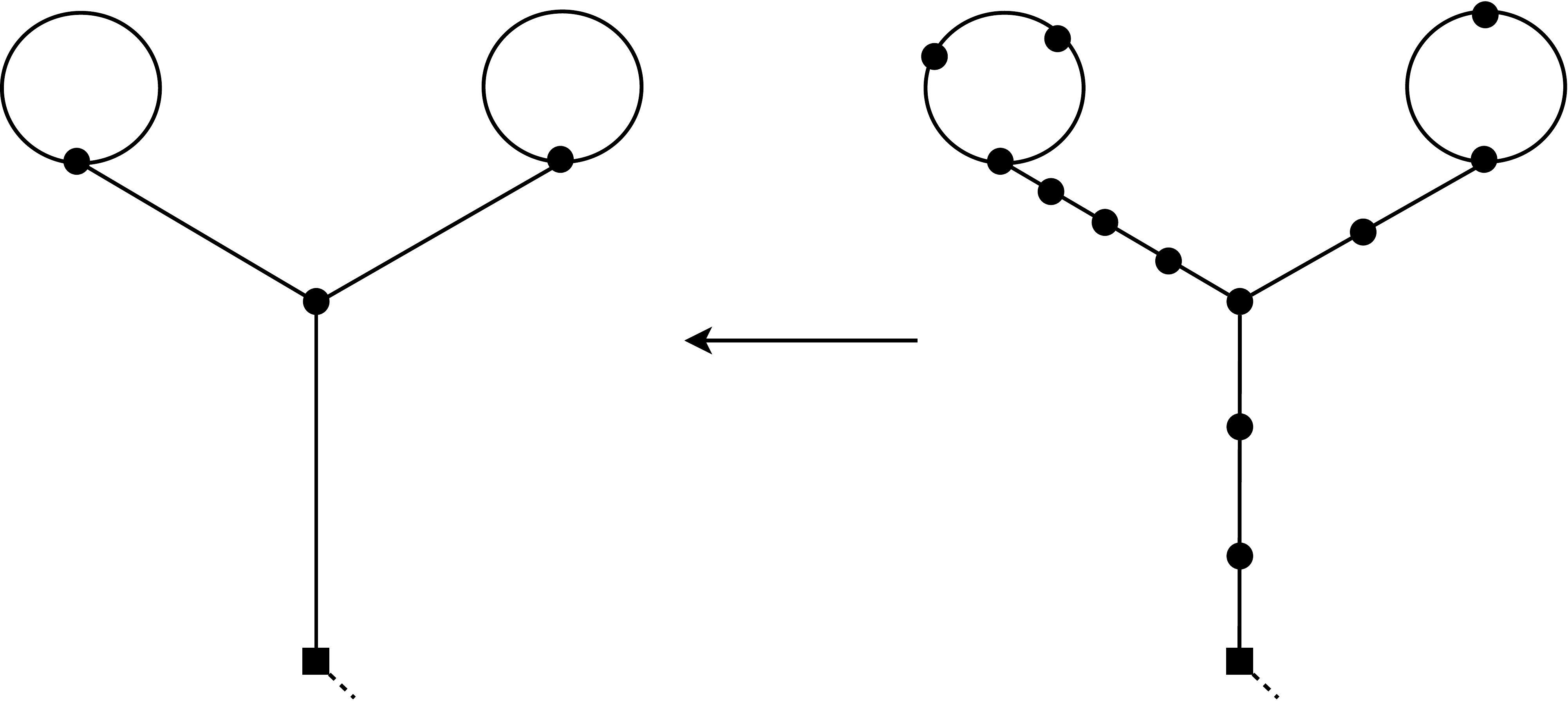}
 \caption{A pruned graph corresponding to the reduced graph $\bar G_1$}
 \label{G12loop}
 \end{center}
\end{figure}

Consider the reduced graph $\bar G_1$ presented in Fig. \ref{G12loop} on the left with both loops having some color $c$ and $c'$, and the 1PR edges multicolored. 
It is a cherry tree. The corresponding $1/N$ contribution is given by summing on the family of pruned graphs presented in Fig. \ref{G12loop} 
on the right and it is 
\bea
\cA (G_1) &&=\frac{1}{N^{2(D-2)}} T^{11}(Dz) \Bigl( \frac{ z}{ (1-zT^2(Dz)) }\Bigr)^2\Bigl(  \frac{ D(D-1) z^2 T^2(Dz)}{[ 1-DzT^2(Dz) ][ 1-zT^2(Dz)  ] }  \Bigr)^3 \crcr
&& =\frac{1}{N^{2(D-2)}}\frac{  D^3(D-1)^3 z^8  T^{14} (Dz)  }{(1-zT^2(Dz))^5 (1-4Dz)^{\frac{3}{2} }  } \sim \frac{1}{N^{2(D-2)}  (1-4Dz)^{\frac{3}{2} }   } \; .
\eea

If one of the two loops is multicolored, the graph $\bar G_1'$ has an extra $N^{-2}$ suppression in $N$ and an extra $\frac{1}{\sqrt{1-4Dz}}$ enhancement factor.
Thus its amplitude is 
\bea
 \cA(\bar G_1') \sim \frac{1}{N^2 \sqrt{1-4Dz}} \cA(\bar G_1) \sim \frac{N^{\frac{D-6}{2}}}{ \sqrt{N^{D-2} (1-4Dz)} } \cA(\bar G_1) \;,
\eea 
hence such reduced graphs are strictly suppressed with respect to $\bar G_1$ in the double scaling limit
$N\to \infty$, $z\to (4D)^{-1}$, $ N^{D-2} [(4D)^{-1}-z]  $ fixed. Furthermore, the reduced graphs having the same structure but without
the vertical $1PR$ edge have less singular factors $\frac{1}{\sqrt{1-4Dz}}$, hence are suppressed. 
 
\begin{figure}[ht]
\begin{center}
\includegraphics[scale=0.25]{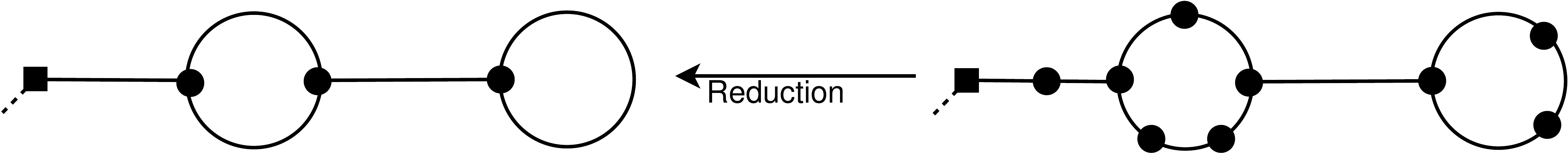}
 \caption{The reduced graph $\bar G_2$ and one pruned graph which projects onto it.}
 \label{G22loop}
 \end{center}
\end{figure}

A second example is the reduced graph $\bar G_2$ of Fig. \ref{G22loop}. 
If its 1PI bars are unicolored, its scaling with $N$ is at best $ N^{-2(D-2)}$ (if the two unicolored edges in the middle have 
the same color)\footnote{The suppression is enhanced to $N^{-2(D-1)}$ if the two edges in the middle have different colors).}
the rest of its amplitude is
\bea
 && T^{11}(Dz) \Bigl( \frac{ z}{ (1-zT^2(Dz)) }\Bigr)^3\Bigl( \frac{ D(D-1) z^2 T^2(Dz)}{[ 1-DzT^2(Dz) ][ 1-zT^2(Dz)  ] }  \Bigr)^2 \crcr
       && = \frac{  z^7 D^2 (D-1)^2 T^{13} (Dz)  }{(1-zT^2(Dz))^3 (1-4Dz)  } \; ,
\eea 
and, as this graph has fewer factors $ \frac{1}{ \sqrt{1-4Dz} }$, it is strictly suppressed in the double scaling regime with respect to $\bar G_1$.

We now consider the case when some of the 1PI edges are multicolored. Note that if the tadpole edge 
is multicolored one always gets a suppressing factor $\frac{1}{ N^2 \sqrt{ 1-4Dz} }$ with respect to the same
reduced graph where the tadpole edge is unicolored (the scaling with $N$ decreases by $N^{-2}$, while only one edge becomes critical).
The most singular case is to have both 1PI edges in the middle multicolored. Such graphs scale like
\bea
 \frac{1}{N^{2D-2}} \frac{1}{ (1-4Dz)^{ 2} } \sim \frac{1}{N^{2(D-2)}  (1-4Dz)^{\frac{3}{2} }   } \; \; \;\frac{1}{N^2 \sqrt{1-4Dz}} \; ,
\eea 
and again are suppressed with respect to $\bar G_1$.

\begin{figure}[ht]
\begin{center}
\includegraphics[scale=0.35]{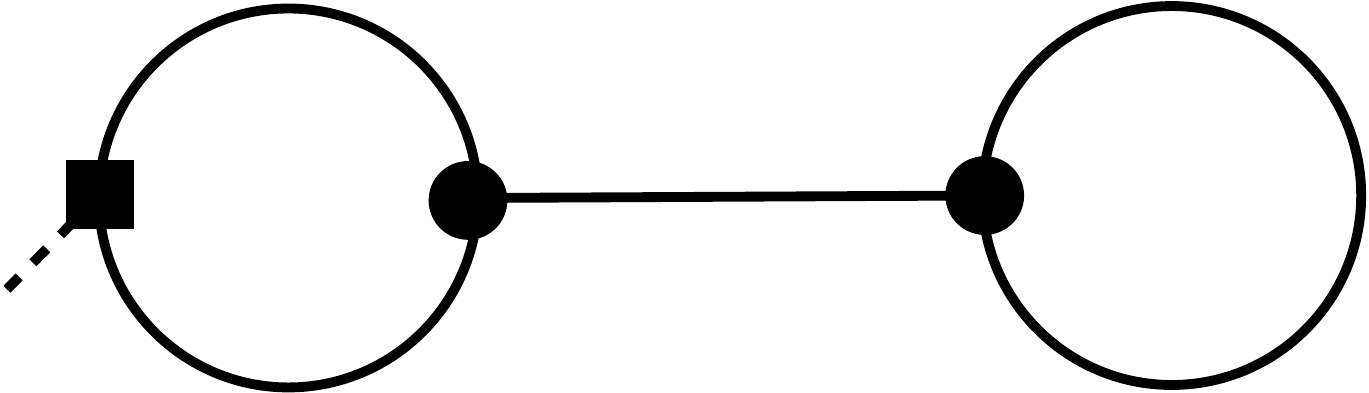}
 \caption{The reduced graph $G_3$.}
 \label{G32loop}
 \end{center}
\end{figure}

Finally our last example is the graph  $\bar G_3$ shown in Fig. \ref{G32loop}. As it has one fewer $1PR$ edge 
than $\bar G_2$, its amplitude is less singular.

\section{Proof of Theorem \ref{maintheo}}

\label{sec:proof}

\subsection{Bounds}

This section is devoted to the proof of \eqref{mainequa} in Theorem \ref{maintheo}.
Recall that by Theorem \ref{thm:ampli}  the amplitude of any reduced graph is 
\bea 
  \cA(\bar G) &=& N^{-D L(\bar G) + 2 \sum_c L( \bar G_c) -2 \sum_c g( \bar G_c)} \; T(Dz)^{1+2E(\bar G)}\\
  &&  \Bigl( \frac{ z }{1-zT^2(Dz)} \Bigr)^{\sum_{c } E( \bar G_c) }
  \Bigl(  \frac{ D(D-1) z^2 T^2(Dz)}{[ 1-DzT^2(Dz) ][ 1-zT^2(Dz)  ] } \Bigr)^{E^{\text{m}}(\bar G)} \; , \nonumber
 \eea 
hence for $z$ close enough to $z_c$ it admits a bound
\bea
 | \cA(\bar G) | \le K_L  N^{-D L(\bar G) + 2 \sum_c L( \bar G_c) -2 \sum_c g( \bar G_c)} (z_c-z)^{-\frac{ E^{\text{m}}(\bar G) }{2}} .
\eea 
Passing to the rescaled variable $x = N^{D-2}(z_c-z)$ this bound writes
\bea
 | \cA(\bar G) | \le K_L  N^{-D L(\bar G) + 2 \sum_c L( \bar G_c) -2 \sum_c g( \bar G_c) + \frac{D-2}{2} E^{\text{m} }  (\bar G)  }
 x^{-\frac{ E^{\text{m}}(\bar G) }{2} } .
\eea 
Using $E(\bar G)\le 3L(\bar G)-1$ and $ E(\bar G_c) = V(G_c) -C(G_c) +L(G_c) $ we have
\bea
  E^{\text{m}}(\bar G) = E(\bar G) - \sum_c E(\bar G_c) \le 3L(\bar G)-1 -\sum_{c}L(G_c)
\eea 
choosing $x<1$ (which includes a neighborhood of $x_c = \frac{1}{4(D-1)}$), 
\bea
 x^{-\frac{ E^{\text{m}}(\bar G) }{2} } \le x^{ - \frac{3}{2}L + \frac{1}{2}} \; .
\eea 
Furthermore, the scaling with $N$ can be bounded as
\bea
&& -D L(\bar G) + 2 \sum_c L( \bar G_c) -2 \sum_c g( \bar G_c) + \frac{D-2}{2} E(\bar G) - \frac{D-2}{2}\sum_c E(\bar G_c) \crcr
&&  \le -\frac{D-2}{2}+\Bigl( -D + \frac{D-2}{2}3\Bigr) L(\bar G) + \Bigl( 2 - \frac{D-2}{2}\Bigr) \sum_c L( \bar G_c) \crcr
&& = -\frac{D-2}{2}- \frac{6-D}{2} L(\bar G) + \frac{6-D}{2} \sum_c L(\bar G_c) \; ,
\eea 
and the bound is saturated only if $E(\bar G) =3L(\bar G) -1$ and $V(\bar G_c) =C(\bar G_c)$ and $g(G_c)=0$. If the bound is not 
saturated then one gets at least a suppressing factor $N^{-\frac{1}{2}}$.
As $D<6$, using the fact that $ \sum_c L(\bar G_c) \le L(\bar G) $, we obtain that the amplitude of any reduced graph is bounded by
\bea 
 A(\bar G) \le K_L N^{\frac{2-D}{2}} x^{-\frac{3}{2}L +\frac{1}{2}}
\eea 
and the scaling in $N$ is saturated only if $E(\bar G) =3L(\bar G) -1$, $V(\bar G_c) =C(\bar G_c)$ and 
$\sum_c L(\bar G_c) = L(\bar G) $, that is $\bar G$ is a cherry tree. For all other graphs one gets at least an extra $N^{-\frac{1}{2}}$ 
suppressing factor, which establishes \eqref{maininequa}.

\subsection{Computation of Cherry Trees}

It remains to prove \eqref{mainequa} by a direct computation.
A full rooted binary tree is a plane tree in which every vertex has either two children or no children. 
The number of full binary trees with $L$ leaves is the Catalan number $C_{L-1}$.
This is also the number of cherry trees with $L$ leaves since a cherry tree is just a full binary tree
plus a single edge from the full binary tree to the ciliated vertex, plus addition of a tadpole with label $c$ on 
each leaf. Hence the map from full binary trees to cherry trees is one to one.

In the special case $L=0$, the reduced graph is just the bare ciliated vertex. It 
corresponds to melons, which we can consider as special degenerate cases of  cherry trees. Hence
\be  \cG^0_{2,cherry}  = T (Dz) .
\ee
For $L \ge 1$ we get from Definition \ref{defcherry} and Theorem \ref{thm:ampli}, taking into account all the possible colorings
of the tadpole edges,
 \bea  \label{equacherry} 
 \cG^L_{2,cherry}(z,N) && = N^{-(D-2)L} C_{L-1} D^{L}  \frac{ z^{ 5 L -2 } D^{2L-1}(D-1)^{2L-1}T^{10L-3}}{(1- zT^2)^{3L-1}  (1- zDT^2 )^{2L-1}  } \crcr
 && =  N^{-(D-2)L} C_{L-1}  \frac{ z^{ 5 L -2 } D^{3L-1}(D-1)^{2L-1}T^{8L-2}}{(1- zT^2)^{3L-1}  (1- 4Dz )^{L-\frac{1}{2}}  } \; .
\eea
In the critical regime $z\to (4D)^{-1}$ we have $T\to 2$, $1-zT^2 \to \frac{D-1}{D}$, $N^{D-2}(1-4Dz) = 4D x $, hence collecting all the factors we get
\bea
  \label{equacherry2} \cG^{L,x}_{2,cherry} (N) =  N^{1-\frac{D}{2}}  \frac{8 \sqrt D C_{L-1} } {[16 (D-1)]^{L} }  x^{-L+\frac{1}{2} } + O(N^{\frac{1}{2}-\frac{D}{2}}) .
\eea 

We recover again the susceptibility of \cite{KOR} by looking at equation \eqref{equacherry} at $L=1$ which give a correction in $N (\lambda - \lambda_c)^{-1}$ to the 
melonic term hence changes $\gamma_0$ from $1/2$ to $3/2$.


\section{The Double Scaling Limit}
\label{sec:Double}

Having completed Theorem \ref{maintheo}
we can forget, for $D<6$, all non cherry trees contributions. We can also forget the $O(N^{1/2-D/2 })$ corrections in the cherry trees themselves
(since they are of same order as the other discarded terms),
keeping their main asymptotic behavior in \eqref{equacherry2} which is
\be  \bar \cG_{2,cherry}^{L,x} (N)  =   N^{1-D/2}  \frac{8\sqrt D  C_{L-1}}{ [16(D-1)]^L}  \;  x^{-L +1/2} .
\ee
Double scaling then consists in summing these contributions over $L$
to find out the leading singularity of the sum at the critical point $x=x_c  = \frac {1}{4(D-1)}$. 

\medskip

The $L=0$ contribution is the melonic contribution $T_{melo}=T(Dz)  $. Adding the sum over cherries gives
the melonic + cherry approximation: 
\bea  \bar \cG^{x}_{2,cherry} (N) &= & T_{melo} +     8  N^{(1-D/2)} \sqrt {Dx}  \sum_{L \ge 1} 
 \frac{ C_{L-1} } {[16 x (D-1)]^{L} } \\
&=&  T_{melo}   +     8  N^{(1-D/2)} \sqrt {Dx}   A\sum_{L \ge 1}  A^{L-1}  C_{L-1} \;,
\eea
 with $A =  [16 x (D-1)]^{-1} $. Hence
\bea
 \bar \cG^{x}_{2,cherry} (N)&=& T_{melo}   +         8  N^{(1-D/2)} \sqrt {Dx}  A   T(A)  \\
&=&  T_{melo}   +    4N^{(1-D/2)}  \sqrt {Dx}  (1-  \sqrt{1- 4A} ) .
\eea
 
Since $\sqrt{1-4A} = x^{-1/2}  \sqrt{x - x_c} $
\bea
\bar \cG^{x}_{2,cherry} (N) &=&  T_{melo}   +    4  N^{(1-D/2)} \sqrt {D} (\sqrt{x}-  \sqrt{x - x_c}  )
\eea

Substituting further $z = z_c - x N^{-D+2}  = \frac{1}{4D} - x N^{-D+2} $ into $T_{melo}$ we can reexpress
\bea
T_{melo} &=&  \frac{2}{1 - 4Dx N^{-D+2}} [1 - 2 \sqrt{DxN^{-D+2}} ] \\
  &=&     2 - 4 N^{(1-D/2)}  \sqrt{Dx}  + O( N^{2-D}),
\eea
and we have obtained 
\begin{theorem} \label{doublelimtheo} The critical point in $x$ is at $x_c = 1 /4(D-1)$ and
the double scaling limit of the quartic  tensor model two point function is
\bea \bar \cG^{x}_{2,cherry} (N) &=&  2 - 4  N^{(1-D/2)} \sqrt {D(x - x_c)}  + O( N^{\frac{1}{2}-\frac{D}{2}}) .
\eea
\end{theorem}

If we rewrite everything in terms of $z$ instead of $x$, we would obtain instead
\bea \label{doublelimbis} \bar \cG^{x}_{2,cherry} (N) &=&  2 - 4  \sqrt {D  [    (z_c -z) -   \frac{N^{2-D}}{4(D-1)}  ] }  +O( N^{\frac{1}{2}-\frac{D}{2}})
\eea
where we used $x = N^{D-2}(z_c - z)$ and $x_c = 1 /4(D-1)$. 

The interpretation of these formulas requires further study. Theorem  \ref{doublelimtheo} shows a square root singularity
in $x-x_c$ of same critical exponent than the melonic singularity. Equation \eqref{doublelimbis} seems to confirm that 
the melonic singularity has slightly moved rather than changed structure.

Triple scaling would consist in writing $(x-x_c) N^{\alpha} =y$ and finding a new correcting expansion in powers of $y$.
It is too early to conclude on a physical interpretation, but the existence of multiple scalings 
suggests that in tensor models of quantum gravity space-time could emerge through a \emph{sequence} of phase transitions,
the first of which corresponds to a  branched polymer phase.

\section*{Acknowledgements}

The authors would like to thank Gilles Schaeffer for introducing them to Wright's approach to graph enumeration.
In particular R.G is indebted to him for numerous discussions on the colored case of \cite{GurauSchaeffer}
which led to the clarification of many of the ideas presented in this paper.

Research at Perimeter Institute is supported by the Government of Canada through Industry Canada and by the Province of
Ontario through the Ministry of Research and Innovation.

 \vskip.5cm
\noindent
{\small ${}^{a}${\it  LIPN, Institut Galil\'ee, CNRS UMR 7030, Universit\'e Paris 13, 
 F-93430, Villetaneuse, France, EU}}
\\
{\small ${}^{b}${\it Laboratoire de Physique Th\'eorique,
Universit\'e Paris 11, 91405 Orsay Cedex, France, EU}} 
\\
{\small ${}^{c}${\it CPHT - UMR 7644, CNRS, \'Ecole Polytechnique, 91128 Palaiseau Cedex, France }}
\\
{\small ${}^{d}${\it Perimeter Institute for Theoretical Physics, 31 Caroline St. N, ON, N2L 2Y5, 
Waterloo, Canada}}\\

\end{document}